\documentclass[a4paper,UKenglish,cleveref, autoref, thm-restate]{lipics-v2021}

\title{The Complexity of X3SAT: P = NP = PSPACE} 

\author{Latif Salum}{Department of Industrial Engineering, Dokuz Eyl\"ul University, Izmir, T\"urkiye}{latif.salum@deu.edu.tr,  latif.salum@gmail.com}{https://orcid.org/0000-0001-5660-1938}{}

\authorrunning{L. Salum}

\Copyright{Latif Salum}



\acknowledgements{I would like to thank Javier Esparza, Anuj Dawar, Avi Wigderson, Paul Spirakis,  \'{E}va Tardos, and Csongor Csehi, as well as every reviewer, for their valuable contributions to the paper throughout its development since 2008. I would like to thank  my colleagues at Dokuz Eyl\"ul University. Last but not least, I would like to thank John Reidar Mathiassen for his sincere efforts in testing the proposed algorithm on real examples.}

\nolinenumbers 

\hideLIPIcs  

\EventEditors{John Q. Open and Joan R. Access}
\EventNoEds{2}
\EventLongTitle{42nd Conference on Very Important Topics (CVIT 2016)}
\EventShortTitle{CVIT 2016}
\EventAcronym{CVIT}
\EventYear{2016}
\EventDate{December 24--27, 2016}
\EventLocation{Little Whinging, United Kingdom}
\EventLogo{}
\SeriesVolume{42}
\ArticleNo{23}

\usepackage{tikz,algorithm,algorithmicx,algpseudocode,mathtools,tabto,xcolor,framed}

\hypersetup{
    colorlinks=true,       
    linkcolor=black!35!cyan,          
    citecolor=black!35!cyan,        
    filecolor=magenta,      
    urlcolor=magenta           
}

\newcommand{\Bullet}{\large $\bullet$}
\newcommand{\cllps}{\tilde{\scp}}
\newcommand{\FM}{\varphi}
\newcommand{\Fm}{\phi}
\newcommand{\fm}{\beta}
\newcommand{\imp}{\Rightarrow}
\newcommand{\Li}{L}
\newcommand{\lt}{r}
\newcommand{\nlt}{\overline{\lt}}
\newcommand{\NP}{\textsf{\bf NP}}
\newcommand{\nx}[1]{\overline{x}_{#1}}
\newcommand{\ny}[1]{\overline{y}_{#1}}
\newcommand{\PP}{\textsf{\bf P}}
\newcommand{\prv}{\vdash}
\newcommand{\rdc}{\rightarrow}
\newcommand{\rdctn}{\tilde{\FM}}
\newcommand{\sat}{\vDash_\alpha\!}
\newcommand{\scp}{\psi}
\newcommand{\shrnk}{\tilde{\Fm}}
\newcommand{\unsat}{\nvDash}
\newcommand{\X}{\odot}

\newcommand{\Cmmnt}[1]{{\small \color{black!50!purple} \Comment{#1}}}

\begin{document}

\maketitle

\begin{abstract}
$C_{k\hskip-0.12em} = (\lt_{i\hskip-0.07em} \X \lt_{j\hskip-0.09em} \X \lt_u)$ is a clause, an exactly-1 disjunction $\X$  of at least \emph{two} literals  $\lt_{i\hskip-0.09em} \in X_{i\hskip-0.05em}$,  $X_{i\hskip-0.11em} = \{x_i, \nx{i}\}$. $C_{k\hskip-0.12em}$ is true if $(\lt_{i\hskip-0.03em} \wedge \nlt_{j\hskip-0.05em} \wedge \nlt_u) \vee (\nlt_{i\hskip-0.03em} \wedge \lt_{j\hskip-0.05em} \wedge \nlt_u) \vee (\nlt_{i\hskip-0.03em} \wedge \nlt_{j\hskip-0.05em} \wedge \lt_u)$ is satisfiable, which leads to  \emph{collapse}, a reduction of $C_{k\hskip-0.05em}$, e.g., $\lt_{i\hskip-0.03em} \wedge C_{k\hskip-0.15em} \prv \nlt_{j\hskip-0.05em} \wedge \nlt_{u\hskip-0.05em}$. Also,   $\nlt_{i\hskip-0.03em} \wedge C_{k\hskip-0.15em} \prv  (\lt_{j\hskip-0.09em} \X \lt_u)$, a  shrinkage. Let $\Fm = \hskip-0.09em\bigwedge\hskip-0.09em C_{k\hskip-0.05em}$, an X3SAT formula. Let $\Fm(\lt_i) \hskip-0.11em\coloneqq \lt_{i\hskip-0.03em} \wedge \Fm$, which leads to  reductions over $\Fm$, viz., $\lt_{i\hskip-0.03em} \wedge \Fm \prv \scp(\lt_i) \wedge \Fm'\hskip-0.05em(\lt_i)$. The  reductions terminate iff $\Li(\scp(\lt_i)) \cap \Li(\Fm'\hskip-0.05em(\lt_i))  = \emptyset$, viz., $\lt_{i\hskip-0.05em} \wedge \Fm \rdc \scp(\lt_i) \wedge \Fm'\hskip-0.05em(\lt_i)$,  in which $\Li(.)  \subseteq  \Li'\hskip-0.09em$, where $\Li'\hskip-0.12em \subseteq \hskip-0.09em \{1, 2, \ldots, n\}$. That is, the reductions  \textbf{terminate} iff $\scp(\lt_i)$ and $\Fm'\hskip-0.05em(\lt_i)$ {\bf are \emph{properly} disjoint} iff neither collapse nor shrinkage  occurs between $\scp(\lt_i)$ and any $C_{k\hskip-0.12em}$ in $\Fm'\hskip-0.05em(\lt_i)$. In this case, unsatisfiability of the formula $\Fm'\hskip-0.05em(\lt_i)$ is \textbf{ignored} to check unsatisfiability of $\Fm(\lt_i)$. Also, $\Fm \supseteq \Fm'\hskip-0.05em(\lt_i)$, and $\scp(\lt_i) =  \lt_{i\hskip-0.05em} \wedge \nlt_{j\hskip-0.07em} \wedge  \cdots \wedge \nlt_{u\hskip-0.05em}$, which is \textbf{consistent}. Otherwise,  $\scp(\lt_j)$ is inconsistent, thus $\Fm(\lt_j)$ is unsatisfiable. Hence, $\lt_{j\hskip-0.12em}$ is removed from $\Fm$ and $j$  from $\Li'\hskip-0.09em$. That is, if $\scp(\lt_j) \prv x_{i\hskip-0.07em} \wedge \nx{i\hskip-0.05em}$, then $\unsat \Fm(\lt_j)$,  $\scp \gets  \scp \wedge \nlt_{j\hskip-0.05em}$, and  $\Li \gets \Li \cup \{j\}$. Next, $\scp \wedge \Fm  \rdc \linebreak \scp^*\hskip-0.15em \wedge \Fm^*\hskip-0.12em$, and  $\Fm^*\hskip-0.07em(\lt_i)$ is re-evaluated for all $i \in \Li'\hskip-0.09em$ and  $\lt_{i\hskip-0.09em} \in X_{i\hskip-0.05em}$, thus $\scp^*\hskip-0.17em \gets \scp^*\hskip-0.15em \wedge \lt_{a\hskip-0.03em}$. $\unsat \Fm$ if $\scp^*\hskip-0.12em$ is inconsistent. Otherwise, $\Fm \rdc \hskip-0.05em\scp \wedge \Fm'\hskip-0.09em$, that is, $\scp$ and $\Fm'\hskip-0.09em$ are properly disjoint, and $\forall i \in \Li' \, \forall \lt_{i\hskip-0.09em} \in X_{i\hskip-0.15em}$ [$\scp(\lt_i)$ is  consistent]. Claim: $\Fm'\hskip-0.12em$ is satisfiable. Proof sketch:  Let  $\Fm \gets \Fm'\hskip-0.09em$. Pick   $\lt_{i_0\hskip-0.15em} \in X_{i_0\hskip-0.19em}$ in $\Fm$, thus $\lt_{i_0\hskip-0.10em} \wedge \Fm \rdc \scp(\lt_{i_0}) \wedge  \Fm'\hskip-0.05em(\lt_{i_0})$. Hence, $\Fm \supseteq \Fm'\hskip-0.05em(\lt_{i_0})$. Pick  $\lt_{i_1\hskip-0.23em} \in X_{i_1\hskip-0.25em}$ in $\Fm'\hskip-0.05em(\lt_{i_0})$, thus $\lt_{i_1\hskip-0.13em} \wedge \Fm'\hskip-0.05em(\lt_{i_0}) \rdc \scp(\lt_{i_1} |\, \lt_{i_0}) \wedge \Fm'\hskip-0.05em(\lt_{i_1} |\, \lt_{i_0})$. That is, $\Fm \rdc \linebreak \scp(\lt_{i_0}) \wedge  \scp(\lt_{i_1} |\, \lt_{i_0}) \wedge \Fm'\hskip-0.05em(\lt_{i_1} |\, \lt_{i_0})$. Also, $\lt_{i_1\hskip-0.09em} \wedge \Fm \rdc  \scp(\lt_{i_1\hskip-0.05em}) \wedge \Fm'\hskip-0.05em(\lt_{i_1\hskip-0.05em})$. Consequently, $\Fm \supseteq \Fm'\hskip-0.05em(\lt_{i_0})$,  $\Fm \rdc  \scp(\lt_{i_1})$ and $\Fm'\hskip-0.05em(\lt_{i_0}) \rdc \scp(\lt_{i_1} |\, \lt_{i_0})$. Thus, $\scp(\lt_{i_1\hskip-0.05em}) \supseteq \scp(\lt_{i_1} |\, \lt_{i_0})$. As $\scp(\lt_{i_1\hskip-0.05em})$ is consistent, $\scp(\lt_{i_1} |\, \lt_{i_0})$ is consistent. Since $\Li(\scp(\lt_{i_0})) \cap \Li(\Fm'\hskip-0.05em(\lt_{i_0}))  = \emptyset$ and $\Fm'\hskip-0.05em(\lt_{i_0}) \rdc \scp(\lt_{i_1} |\, \lt_{i_0})$, $\scp(\lt_{i_0})$ and $\scp(\lt_{i_1} |\, \lt_{i_0})$ are \emph{properly} disjoint. Then, $\scp(\lt_{i_1} |\, \lt_{i_0})$ can be appended to $\scp(\lt_{i_0})$, i.e., $\big(\scp(\lt_{i_0}) \wedge  \scp(\lt_{i_1} |\, \lt_{i_0})\big)$ is consistent. Also, $\scp(\lt_{i_2} |\, \lt_{i_1})$ can be appended to $\big(\scp(\lt_{i_0}) \wedge  \scp(\lt_{i_1} |\, \lt_{i_0})\big)$. Thus, $\hat{\scp} = \scp(\lt_{i_0}) \wedge \bigwedge^n_{k = 1} \hskip-0.11em \scp(\lt_{i_k} |\, \lt_{i_{k-1}})$, which is \emph{consistent}. That is, construction of the next  $\scp(\lt_{i_k} |\, \lt_{i_{k-1}})$ depends \emph{only} upon the current  $\Fm'\hskip-0.05em(\lt_{i_{k-1}} |\, \lt_{i_{k-2}})$. Thus, $\Fm'\hskip-0.12em \rdc \hat{\scp}$, and $\Fm'\hskip-0.12em$ is \emph{satisfiable}. To tackle \textit{TQBF},  the Prime Normal Form $\Psi$ is constructed over a 3SAT $\fm$, in which $\fm =  \hskip-0.09em \bigwedge^m_{k=1} \hskip-0.07em c_{k\hskip-0.09em}$ and $\Psi =  \bigwedge \hskip-0.09em \delta_{\hskip-0.05em k\hskip-0.05em}$,  where $\delta_{\hskip-0.05em k\hskip-0.11em} = (\scp_k^{_1\!} \vee \scp_k^{_2\!} \vee \cdots \vee \scp_k^{_7})$ such that $\scp_k^i \hskip-0.05em  \wedge \fm$ is satisfiable. Also, $\scp_k^{_1\!} \hskip-0.07em= \lt_{i\hskip-0.05em} \wedge \nlt_{j\hskip-0.09em} \wedge \nlt_{u\hskip-0.05em}$, $\scp_k^{_2\!} \hskip-0.07em= \nlt_{i\hskip-0.05em} \wedge \lt_{j\hskip-0.09em} \wedge \nlt_{u\hskip-0.05em}, \ldots, \scp_k^{_7\!} \hskip-0.07em= \lt_{i\hskip-0.05em} \wedge \lt_{j\hskip-0.09em} \wedge \lt_{u\hskip-0.05em}$, which denote the prime satisfying assignments for the clause $c_{k\hskip-0.12em} = (\lt_{i\hskip-0.11em} \vee \lt_{j\hskip-0.15em} \vee \lt_u)$. The complexity of \textit{X3SAT} is $O(mn^3)$, and of \textit{TQBF}  is $O(m^2n^3)$. Thus, $\PP = \NP = \textsf{\bf PSPACE}$. The paper also tackles Graph Isomorphism via \textit{XSAT}.
\end{abstract}

\section{Introduction}
If any \NP-complete problem is in $\PP$, then $\PP = \NP$. In this respect, there is no difference in proving that \textit{3SAT} is in $\PP$ and proving that \textit{CLIQUE} is in $\PP$. Nevertheless, a particular   problem  may feature a  property that leads to an efficient algorithm, which proves  $\PP = \NP$.

This paper shows  $\PP = \NP$ via {\it One-in-Three 3SAT}, also called {\it Exactly-1 3SAT} or \textit{X3SAT}, which is \NP-complete \cite{Sch78}. \textit{X3SAT} features X\.{O}R (exactly-1 or), denoted by $\X$. X\.{O}R leads to  an efficient algorithm that decides  satisfiability of an X3SAT formula $\Fm$.  The algorithm, called the (formula) $\Fm$ scan, incorporates a proof theoretic approach.  The following introduces the $\Fm$ scan. See also this \href{https://arxiv.org/abs/2203.04142}{reply} and  \href{https://www.academia.edu/92564031/Tractability_of_Exactly_1_3SAT_P_NP}{presentation}.

$C_{k\hskip-0.12em} = (\lt_{i\hskip-0.07em} \X \lt_{j\hskip-0.09em} \X \lt_u)$ denotes a clause,   which involves at least \emph{two} literals $\lt_{i\hskip-0.05em}$,   $\lt_{i\hskip-0.09em} \in \{x_i, \nx{i}\}$. $C_{k\hskip-0.09em}$ is true iff \emph{exactly one} of $\{\lt_i, \lt_j, \lt_u\}$ is  true. Then, $\Fm = \hskip-0.09em\bigwedge \hskip-0.09em C_{k\hskip-0.12em}$ denotes an X3SAT formula.

Let $\Fm(\lt_j) \hskip-0.09em\coloneqq \lt_{j\hskip-0.09em} \wedge \Fm$ for any $j \in \Li'\hskip-0.09em$,  $\Li'\hskip-0.12em \subseteq \{1, 2, \ldots, n\}$. If $\Fm(\lt_j)$ is unsatisfiable, viz., $\unsat \Fm(\lt_j)$, then $\lt_{j\hskip-0.12em}$ is incompatible. Consider $\Fm(x_j)$. Then, each   $C_{k\hskip-0.09em}$ containing $x_{j\hskip-0.09em}$ or $\nx{j\hskip-0.12em}$ is reducible, viz., $x_{j\hskip-0.09em} \wedge (x_{j\hskip-0.07em} \X \nx{i\hskip-0.07em} \X x_u) \wedge (\nx{j\hskip-0.07em} \X \nx{u\hskip-0.09em} \X x_v) \prv  x_{i\hskip-0.07em} \wedge \nx{u\hskip-0.07em} \wedge (\nx{u\hskip-0.09em} \X x_v)$. Hence, $\nx{u\hskip-0.12em}$ leads to the subsequent reduction, i.e., $\nx{u\hskip-0.07em} \wedge (\nx{u\hskip-0.09em} \X x_v) \prv \nx{v\hskip-0.05em}$. Thus, $x_{j\hskip-0.09em} \wedge \Fm$ is reduced to $\scp(x_j) \wedge \Fm'\hskip-0.05em(x_j)$, i.e., $x_{j\hskip-0.07em} \wedge \Fm \prv \linebreak \scp(x_j) \wedge \Fm'\hskip-0.05em(x_j)$, where $\scp(x_j) =  x_{j\hskip-0.09em} \wedge x_{i\hskip-0.07em} \wedge \nx{u\hskip-0.07em} \wedge \nx{v\hskip-0.05em}$. Note that $\scp(x_j)$ is a conjunction of literals, called a minterm. Next, $x_{i\hskip-0.07em}$ and $\nx{v\hskip-0.07em}$ proceed the reductions over $\Fm'\hskip-0.05em(x_j)$, and $\scp(x_j) \hskip-0.07em\gets \scp(x_j) \wedge \lt_{a\hskip-0.03em}$.

If $\scp(x_j)$ is inconsistent,  viz., $\scp(x_j) \prv x_{i\hskip-0.07em} \wedge \nx{i\hskip-0.12em}$ for some $i$, then $\unsat \Fm(x_j)$, hence $x_{j\hskip-0.12em}$ is  removed from $\Fm$ and $j$  from $\Li'\hskip-0.09em$, viz., $\scp \gets \scp \wedge \nx{j\hskip-0.09em}$ and $\Li \gets \Li \cup \{j\}$. Otherwise,  $x_{j\hskip-0.07em} \wedge \Fm \rdc \scp(x_j) \wedge \Fm'\hskip-0.05em(x_j)$. In this case, unsatisfiability of $\Fm'\hskip-0.05em(x_j)$ is \emph{ignored} to check unsatisfiability of $\Fm(x_j)$. Also,  $\scp(x_j)$ and $\Fm'\hskip-0.05em(x_j)$ become \emph{properly} disjoint. That is, if $i \in \Li(\scp(x_j))$, then $X_{i\hskip-0.07em} \cap C_{k\hskip-0.12em} = \emptyset$ for any $C_{k\hskip-0.12em}$ in $\Fm'\hskip-0.05em(x_j)$, and if $i \in \Li(\Fm'\hskip-0.05em(x_j))$, then $\scp(x_j) \cap X_{i\hskip-0.11em}  = \emptyset$, where $X_{i\hskip-0.09em} = \{x_i, \nx{i}\}$ and $\Li(.) \subseteq \Li'\hskip-0.09em$.

Next, $\scp \wedge \Fm \rdc \scp^*\hskip-0.17em \wedge \Fm^*\hskip-0.12em$. Then,  $\Fm^*\hskip-0.07em(\lt_i)$ is re-evaluated for all $i \in \Li'\hskip-0.09em$ and  $\lt_{i\hskip-0.09em} \in X_{i\hskip-0.05em}$. Thus, $\scp^*\hskip-0.19em \gets \linebreak \scp^*\hskip-0.12em \wedge \lt_{a\hskip-0.03em}$, hence $\unsat \Fm$ if $\scp^*\hskip-0.12em$ is inconsistent. Otherwise, the $\Fm$ scan terminates, viz., $\Fm \rdc \scp \wedge \Fm'\hskip-0.12em$ (see \cref{f:Scan}). That is, $\scp$ and $\Fm'\hskip-0.09em$ are \emph{properly} disjoint, and   \color{blue}$\forall i \in \Li' \, \forall \lt_{i\hskip-0.09em} \in X_{i\hskip-0.15em}$ [$\scp(\lt_i)$ is  consistent]\color{black}.

\begin{figure} [!h]
\begin{framed}
\centering
\vskip-0.7em

\small
$\unsat \Fm(x_5)$ if $\scp(x_5)$ is inconsistent. $x_{5\hskip-0.09em}$ is incompatible and to be removed from $\Fm$.   $\scp \gets \scp \wedge \nx{5\hskip-0.05em}$.

Let $\scp(x_3)$ be consistent, i.e., $x_{3\hskip-0.05em} \wedge \Fm \rdc \scp(x_3) \wedge \Fm'\hskip-0.05em(x_3)$.   Then, unsatisfiability of $\Fm'\hskip-0.05em(x_3)$ is \emph{ignored}.

$\scp \gets \scp \wedge x_{4\hskip-0.09em}$ if $\scp(\nx4)$ is inconsistent. As a result, $\Fm(x_3)$ is re-evaluated.

$\scp \gets \scp \wedge \nx{3\hskip-0.09em}$ if $\scp(x_3)$ is inconsistent. $\scp = \nx{5\hskip-0.03em} \wedge x_{4\hskip-0.05em} \wedge \nx{3\hskip-0.03em}$, i.e., $\Li = \{3, 4, 5\}$. $\Li'\hskip-0.15em = \{1, 2, 6, 7, 8, 9\}$.

\vskip0.4em

$\underset{\displaystyle \scp}{\underbrace{\nx{5\hskip-0.03em} \wedge x_{4\hskip-0.05em} \wedge \nx3}} \wedge
  \underset{\displaystyle \Fm}
   {\underbrace{
        C_{1\hskip-0.09em} \wedge \cdots \wedge
                \overset{\displaystyle \scp(x_3) \prv x_{3\hskip-0.07em} \wedge {\color{red}\nx{8\hskip-0.07em}} \wedge \nx{1\hskip-0.09em} \wedge x_{6\hskip-0.07em} \wedge \nx{9\hskip-0.09em}  \wedge {\color{red}x_{8\hskip-0.07em}}}
                {\overbracket[0.6pt]{
        (x_{3\hskip-0.07em} \X x_{8\hskip-0.07em} \X x_1) \wedge (x_{3\hskip-0.07em} \X \nx{6\hskip-0.07em} \X x_9) \wedge (\nx{6\hskip-0.07em} \X x_8)
                }} \wedge \cdots \wedge C_m
    }}$.

\vskip-0.1em
\begin{tikzpicture} [transform shape, scale=0.85]
    \draw[|-|, thick, dashed] +(-0.5,0)  node[left] {\Large$\Fm'\hskip-0.09em$} to (12.5,0);

    \draw[|-|] (-0.5  ,0.4)  to node[above] {$\color{blue}\scp(x_2)$}   (1.7,0.4);    \draw[|-|] (1.4,0.2)   to node[above] {$\color{blue}\scp(\nx1)$} (4.5,0.2);
    \draw[|-|] (4.3,0.4)     to node[above] {$\color{blue}\scp(\nx6)$}  (6.5  ,0.4);  \draw[|-|] (6,0.2)     to node[above] {$\color{blue}\scp(x_7)$}  (9.5 ,0.2);
    \draw[|-|] (8.5  ,0.4)   to node[above] {$\color{blue}\scp(x_8)$}   (11 ,0.4);    \draw[|-|] (10.8,0.2)  to node[above] {$\color{blue}\scp(\nx9)$} (12.5 ,0.2);

    \draw[|-|] (-0.5  ,-0.4) to node[below] {$\color{blue}\scp(x_6)$}   (1.3,-0.4);   \draw[|-|] (0.7,-0.2)  to node[below] {$\color{blue}\scp(\nx2)$} (3.9,-0.2);
    \draw[|-|] (3.4,-0.4)    to node[below] {$\color{blue}\scp(\nx7)$}  (5.5  ,-0.4); \draw[|-|] (5,-0.2)    to node[below] {$\color{blue}\scp(\nx8)$} (9,-0.2);
    \draw[|-|] (8,-0.4)      to node[below] {$\color{blue}\scp(x_1)$}   (10.7 ,-0.4); \draw[|-|] (10.5,-0.2) to node[below] {$\color{blue}\scp(x_9)$}  (12.5,-0.2);
 \end{tikzpicture}\vskip-1.2em
\end{framed}\vskip-0.3em

  \caption{$\Fm \rdc \scp \wedge \Fm'\hskip-0.09em$. $n = 9$, $\scp = \nx{5\hskip-0.03em} \wedge x_{4\hskip-0.05em} \wedge \nx{3\hskip-0.03em}$, and $\Fm'\hskip-0.19em =  C_{1\hskip-0.09em} \wedge \cdots \wedge (x_{8\hskip-0.07em} \X x_1) \wedge (\nx{6\hskip-0.07em} \X x_9) \wedge \cdots \wedge C_{m\hskip-0.05em}$.}\label{f:Scan}
\end{figure}
\vspace{-0.6em}

\begin{claim*}
$\Fm$ is satisfiable iff $\Fm \rdc \scp \wedge \Fm'\hskip-0.09em$, that is, $\Fm$ is satisfiable iff the $\Fm$ scan terminates.
\end{claim*}
\begin{claimproof}[Proof sketch]
Since the $\Fm$ scan terminates, $\scp(\lt_i)$ is  consistent for any $i \in \Li'\hskip-0.09em$ and  $\lt_{i\hskip-0.09em} \in X_{i\hskip-0.05em}$. Let  $\Fm \gets \Fm'\hskip-0.09em$. Then, $\lt_{i_0\hskip-0.11em} \wedge \Fm \rdc \color{blue}\scp(\lt_{i_0})\color{black} \wedge \Fm'\hskip-0.05em(\lt_{i_0}), \, \lt_{i_1\hskip-0.17em} \wedge \Fm'\hskip-0.05em(\lt_{i_0})  \rdc \color{blue}\scp(\lt_{i_1} |\, \lt_{i_0})\color{black} \wedge \Fm'\hskip-0.05em(\lt_{i_1} |\, \lt_{i_0}), \ldots,  \lt_{i_{n-1\hskip-0.11em}} \wedge \Fm'\hskip-0.05em(\lt_{i_{n-2}} |\, \lt_{i_{n-3}})  \rdc \color{blue}\scp(\lt_{i_{n-1}} |\, \lt_{i_{n-2}})\color{black} \wedge \Fm'\hskip-0.05em(\lt_{i_{n-1}} |\, \lt_{i_{n-2}}), \,\lt_{i_n\hskip-0.15em} \wedge  \Fm'\hskip-0.05em(\lt_{i_{n-1}} |\, \lt_{i_{n-2}}) \rdc  \color{blue}\scp(\lt_{i_n} |\, \lt_{i_{n-1}})$, thus $\Li(\scp(\lt_{i_0})) \cap \Li(\Fm'\hskip-0.05em(\lt_{i_0}))  = \emptyset$, and $\Li(\scp(\lt_{i_{k-1}} |\, \lt_{i_{k-2}})) \cap \Li(\Fm'\hskip-0.05em(\lt_{i_{k-1}} |\, \lt_{i_{k-2}}))  = \emptyset$. Hence, $\scp(\lt_{i_0})$ and $\scp(\lt_{i_1} |\, \lt_{i_0})$ are \emph{properly} disjoint by   $\Fm'\hskip-0.05em(\lt_{i_0}) \rdc \scp(\lt_{i_1} |\, \lt_{i_0})$, and $\scp(\lt_{i_{k-1}} |\, \lt_{i_{k-2}})$ and $\scp(\lt_{i_k} |\, \lt_{i_{k-1}})$ are \emph{properly} disjoint by $\Fm'\hskip-0.05em(\lt_{i_{k-1}} |\, \lt_{i_{k-2}})  \rdc \scp(\lt_{i_k} |\, \lt_{i_{k-1}})$. Also, $\scp(\lt_{i_k} |\, \lt_{i_{k-1}})$ is consistent, since $\scp(\lt_{i_k}) \supseteq \scp(\lt_{i_k} |\, \lt_{i_{k-1}})$. Thus, $\Fm'\hskip-0.12em \rdc \color{blue} \scp(\lt_{i_0}) \wedge  \scp(\lt_{i_1} |\, \lt_{i_0}) \wedge \cdots \wedge \scp(\lt_{i_{n-1}} |\, \lt_{i_{n-2}}) \wedge \scp(\lt_{i_n} |\, \lt_{i_{n-1}})$. That is, $\Fm'\hskip-0.09em$  is reducible to a minterm  \emph{consistent}. Therefore, $\Fm'\hskip-0.09em$, hence $\Fm$, is \emph{satisfiable}. See also \cref{f:OvCntn}, in which $\Fm = (x_{1\hskip-0.12em} \X \nx{2\hskip-0.05em} \X x_9) \wedge (x_{7\hskip-0.05em} \X \nx{2\hskip-0.05em} \X \nx8) \wedge (\nx{6\hskip-0.05em} \X x_9)$, $\Fm \subseteq \Fm'\hskip-0.09em$ from \cref{f:Scan}.
\end{claimproof}

\begin{figure} [!h]
\centering
\vskip-1.1em
  \begin{tikzpicture}[transform shape, scale=0.9]
     \draw[|-|]                +(0,0)        node[left]  {$x_{1\hskip-0.11em} \wedge \Fm \rdc \scp(x_1) \wedge \Fm'\hskip-0.05em(x_1)$}                                 to  (8,0);
      \path                     (0,0)     to node[above]
         {$\scp(x_1) = x_{1\hskip-0.15em} \wedge  x_{2\hskip-0.09em} \wedge \nx{9\hskip-0.09em} \wedge \nx{6\hskip-0.03em}$, defined over $\Fm$}                            (8,0);
     \draw[-|, thick, dashed]   (8,0)     to node[above] {$\Fm'\hskip-0.05em(x_1) = (x_{7\hskip-0.05em} \X \nx8)$}                                                          (11.5,0);

    \draw[|-|]                 +(0,-0.8)     node[left]  {$x_{2\hskip-0.07em} \wedge \Fm \rdc \scp(x_2) \wedge \Fm'\hskip-0.05em(x_2)$}                                 to  (2,-0.8);
     \path                      (0,-0.8)  to node[above] {$\scp(x_2) = x_2$}                                                                                                (2,-0.8);
      \draw[-|, thick, dashed]  (2,-0.8)  to node[above]
         {$\Fm'\hskip-0.05em(x_2) = (x_{1\hskip-0.12em} \X  x_9) \wedge (x_{7\hskip-0.05em} \X \nx8) \wedge (\nx{6\hskip-0.05em} \X x_9)$, contained by $\Fm$}              (11.5,-0.8);

    \draw[|-|]               +(2,-1.6)     node[left]  {$x_{1\hskip-0.11em} \wedge \Fm'\hskip-0.05em(x_2) \rdc \scp(x_1 |\, x_2) \wedge \Fm'\hskip-0.05em(x_1 |\, x_2)$} to (8,-1.6);
     \path                    (2,-1.6)  to node[above]
         {$\scp(x_1 |\, x_2) = x_{1\hskip-0.15em} \wedge \nx{9\hskip-0.09em} \wedge \nx{6\hskip-0.03em}$, over $\Fm'\hskip-0.05em(x_2)$}
                                           node[below]
         {\small $\scp(x_1) \supseteq \scp(x_1 |\, x_2)$}                                                                                                                          (8,-1.6);
      \draw[-|, thick, dashed]  (8,-1.6)  to node[above] {$\Fm'\hskip-0.05em(x_1 |\, x_2) = (x_{7\hskip-0.05em} \X \nx8)$}                                                  (11.5,-1.6);
  \end{tikzpicture}

  \caption{$\Fm \rdc  \scp(x_1) \wedge \scp(x_7 \mspace{1mu}|\, x_1)$, where $\scp(x_7 \mspace{1mu}|\, x_1) = x_{7\hskip-0.07em} \wedge x_{8\hskip-0.07em}$ due to $x_{7\hskip-0.07em} \wedge \Fm'\hskip-0.05em(x_1) \rdc \scp(x_7 \mspace{1mu}|\, x_1)$. Also, $\Fm \rdc \scp(x_2) \wedge \scp(x_{1\hskip-0.07em} \mspace{1mu}|\, x_2) \wedge \scp(\nx7 \mspace{1mu}|\, x_1)$, where $\scp(\nx7 \mspace{1mu}|\, x_1) = \nx{7\hskip-0.09em} \wedge \nx{8\hskip-0.07em}$ due to $\nx{7\hskip-0.09em} \wedge \Fm'\hskip-0.05em(x_1 \mspace{1mu}|\, x_2) \rdc \scp(\nx7 \mspace{1mu}|\, x_1)$.}\label{f:OvCntn}
  \vskip-1.5em
\end{figure}

\section{Basic Definitions}
\textit{X3SAT} features X\.{O}R, which facilitates deciding unsatisfiability. A formula is \emph{unsatisfiable} if it is reducible to a simple formula \emph{inconsistent}, viz., $\unsat \Fm$ if $\Fm \prv \scp$ such that $\scp  \prv x_{i\hskip-0.07em} \wedge \nx{i\hskip-0.05em}$.

\begin{definition}[Literal]\label{literal}
 $\lt_{i\hskip-0.12em}$ denotes a Boolean variable $x_{i\hskip-0.12em}$ or its negation $\nx{i\hskip-0.05em}$, that is, $\lt_{i\hskip-0.09em} \in X_{i\hskip-0.12em}$ for any $i \in \mathfrak{\Li}$, in which $X_{i\hskip-0.11em} = \{x_i, \nx{i}\}$ and  $\mathfrak{\Li} = \{1, 2, \ldots, n\}$.
\end{definition}

\begin{definition}[X\.{O}R]\label{Xdsn}
$\lt_{i\hskip-0.07em} \X \lt_{j\hskip-0.09em} \X \cdots \X \lt_{u\hskip-0.17em}$ is true iff exactly one of $\{\lt_i, \lt_j, \ldots, \lt_u\}$ is true iff $\delta$ is satisfiable, in which $\delta = \dot{\scp}(\lt_i) \vee \dot{\scp}(\lt_j) \vee \cdots \vee \dot{\scp}(\lt_u)$, where  $\dot{\scp}(\lt_i) = \lt_{i\hskip-0.07em} \wedge \nlt_{j\hskip-0.09em} \wedge \cdots \wedge \nlt_{u\hskip-0.05em}$.
\end{definition}

\begin{definition}[Clause]\label{clause}
$C_{k\hskip-0.12em} = (\lt_{i\hskip-0.07em} \X \lt_{j\hskip-0.09em} \X \cdots \X \lt_u)$, i.e., $C_{k\hskip-0.12em} = \{\lt_i, \lt_j, \ldots, \lt_u\}$, for any $k \in \mathfrak{C}$, where $\mathfrak{C} = \{1, 2, \ldots, m\}$. Also, $|C_k| \in \{2, 3\}$ for X3SAT and $|C_k| \in \{2, 3, \ldots, n\}$ for XSAT.
\end{definition}

\begin{definition}[Minterm/Simple formula]\label{minterm}
$\scp = \lt_{i\hskip-0.05em} \wedge \lt_{j\hskip-0.07em} \wedge \cdots \wedge \lt_{v\hskip-0.03em}$, i.e., $\scp = \{\lt_i, \lt_j, \ldots, \lt_v\}$, in which a literal  denotes a \emph{conjunct}. Any conjunct is \emph{necessary} for satisfying  some formula.
\end{definition}

\begin{definition}\label{incnstnt}
A simple formula $\scp$ is inconsistent iff $\scp  \prv x_{i\hskip-0.07em} \wedge \nx{i\hskip-0.09em}$ for some $i$.
\end{definition}

\begin{definition}[Initial X3SAT formula]\label{formula}
$\FM =  \scp \wedge \Fm$, where $\Fm = \hskip-0.09em\bigwedge_{k \in \mathfrak{C}} C_{k\hskip-0.12em}$ and $|C_k| \in \{2, 3\}$.
\end{definition}

\begin{definition}\label{LL'}
$\mathfrak{\Li} = \Li \cup \Li'\hskip-0.09em$, where $\Li = \{ j \mid \lt_{j\hskip-0.12em} \in \scp\}$  and  $\Li'\hskip-0.15em = \{ i \mid \lt_{i\hskip-0.09em} \in C_{k\hskip-0.12em} \text{ for some } C_{k\hskip-0.09em} \text { in } \Fm\}$.
\end{definition}

\begin{note*}
$\Li \cap \Li'\hskip-0.15em \neq \emptyset$, if $\scp \neq \emptyset$. $\Li \cap \Li'\hskip-0.15em = \emptyset$, \emph{whenever} the reductions due to $\scp$ over $\Fm$ terminate.
\end{note*}

\begin{definition}\label{ChkIncmp}
$\Fm(\lt_i) = \lt_{i\hskip-0.05em} \wedge \Fm$ for any $i \in \Li'\hskip-0.12em$ and $\lt_{i\hskip-0.11em} \in X_{i\hskip-0.05em}$, $X_{i\hskip-0.11em} = \{x_i, \nx{i}\}$, \emph{whenever}  $\Li \cap \Li'\hskip-0.15em = \emptyset$.
\end{definition}

\begin{definition}\label{TrvIncmp}
If $\nlt_{j\hskip-0.15em}$ is necessary, then $\lt_{j\hskip-0.17em}$ is incompatible \emph{trivially}, and removed, viz., $\nlt_{j\hskip-0.09em} \prv \neg \lt_{j\hskip-0.07em}$.
\end{definition}

\begin{definition}\label{NonTrvIncmp}
If $\Fm(\lt_j)$ is unsatisfiable, then $\lt_{j\hskip-0.17em}$ is incompatible \emph{nontrivially}. As a result, it is removed from $\Fm$, thus  $\nlt_{j\hskip-0.17em}$ is necessary for $\FM$, viz., if $\unsat \Fm(\lt_j)$, then $\neg \lt_{j\hskip-0.09em} \prv \nlt_{j\hskip-0.05em}$, thus $\scp \gets \scp \wedge \nlt_{j\hskip-0.05em}$.
\end{definition}

\begin{definition}\label{PrpDsj}
The sets $A$ and $B$ are  properly disjoint with respect to  $X_{i\hskip-0.11em}$ iff $A \cap X_{i\hskip-0.11em}  = \emptyset$ for any $i \in \Li(B)$ and $X_{i\hskip-0.09em} \cap B = \emptyset$ for any $i \in \Li(A)$, where $X_{i\hskip-0.11em} = \{x_i, \nx{i}\}$ and $\Li(.) \subseteq \mathfrak{\Li}$.
\end{definition}

\begin{lemma}[Collapse of a clause to a minterm]\label{cllps}
$\lt_{i\hskip-0.07em} \wedge C_{k\hskip-0.09em} \prv \scp_k(\lt_i)$, thus $C_{k\hskip-0.12em}$ becomes empty, in which $\scp_k(\lt_i) = \nlt_{j\hskip-0.07em} \wedge \cdots \wedge \nlt_{u\hskip-0.09em}$ for $C_{k\hskip-0.12em} = (\lt_{i\hskip-0.07em} \X \lt_{j\hskip-0.09em} \X \cdots \X \lt_u)$, where $i \neq j, \ldots, i \neq u$.
\end{lemma}
\begin{proof}
Follows directly from   \cref{Xdsn}. Note that  $\lt_{i\hskip-0.07em} \wedge C_{k\hskip-0.12em}$ is true iff $\dot{\scp}_k(\lt_i)$ is true.
\end{proof}

\begin{lemma}[Shrinkage of a clause]\label{shrnk}
$\nlt_{j\hskip-0.07em} \wedge C_{k\hskip-0.09em}  \prv C_k(\neg \lt_j)$ such that if $C_k(\neg \lt_j) = (\lt_u)$, then $(\lt_u)$ the unit clause becomes $\lt_{u\hskip-0.15em}$ the \emph{conjunct}, that is, $\lt_{u\hskip-0.15em}$ becomes \emph{necessary} for $\FM$, or for $\Fm(\lt_i)$.
\end{lemma}
\begin{proof}
Follows from  \cref{clause,TrvIncmp,minterm}. Note that  $C_{k\hskip-0.09em}$ contains at least \emph{two} literals.
\end{proof}

\begin{note}\label{rdctn}
Collapse (or shrinkage) denotes a reduction, a syntactic consequence. A reduction arises \emph{firstly} by \cref{formula},  $\scp \wedge \Fm \prv \scp'\hskip-0.12em \wedge \Fm'\hskip-0.09em$, or \emph{secondly} by \cref{ChkIncmp},   $\lt_{i\hskip-0.05em} \wedge \Fm \prv \scp(\lt_i) \wedge \Fm'\hskip-0.05em(\lt_i)$.
\end{note}

\begin{remark}\label{rdctnTrm}
Reductions over $\Fm$  are denoted by $\prv$, and their termination is denoted by $\rdc$.
\end{remark}

\begin{example}
Let $\FM = \nx{1\hskip-0.07em} \wedge (x_{1\hskip-0.12em} \X \nx{2\hskip-0.04em} \X x_3) \wedge (\nx{3\hskip-0.05em}  \X \nx4) \wedge (\nx{3\hskip-0.05em} \X \nx{2\hskip-0.04em} \X x_1)$. Hence, $\FM \rdc  \Fm'\hskip-0.09em$, where $\Fm'\hskip-0.15em =  (\nx{2\hskip-0.04em} \X x_3) \wedge (\nx{3\hskip-0.05em}  \X \nx4) \wedge (\nx{3\hskip-0.05em} \X \nx{2})$ (\cref{rdctn}, the first case). Thus, $\Li = \{1\}$ and $\Li'\hskip-0.15em = \{2, 3, 4\}$ by \cref{LL'}. Let $\Fm \gets \Fm'\hskip-0.09em$. Consider $\Fm(x_4)$ by \cref{ChkIncmp}. Then, $x_{4\hskip-0.03em} \wedge \Fm \prv (\nx{2\hskip-0.07em} \X x_3) \wedge  \nx{3\hskip-0.03em} \wedge \linebreak (\nx{3\hskip-0.07em} \X \nx2) \prv   \nx{2\hskip-0.03em} \wedge x_{2\hskip-0.07em}$ (\cref{rdctn}, the second case). Hence,  $\unsat \Fm(x_4)$, thus $\neg x_{4\hskip-0.09em} \prv  \nx{4\hskip-0.05em}$ and $\scp \gets  \scp \wedge  \nx{4\hskip-0.05em}$ by \cref{NonTrvIncmp}. Also, $\unsat \Fm(\nx4)$ and $\scp \gets \scp \wedge x_{4\hskip-0.03em}$. That is, $\unsat \Fm$, because $\Fm \prv \scp$ and $\scp \prv \nx{4\hskip-0.03em} \wedge x_{4\hskip-0.05em}$.
\end{example}

Reductions  underlie the $\FM$ scan,  the algorithm that decides  satisfiability of  a formula  $\FM$ (see \cref{formula}).  Consider the formula $\Fm(\lt_i)$ by \cref{ChkIncmp}. Then, the reductions transform $\lt_{i\hskip-0.05em} \wedge \Fm$ into $\scp(\lt_i) \wedge \Fm'\hskip-0.05em(\lt_i)$, unless $\scp(\lt_i)$ is inconsistent, such that  $\scp(\lt_i)$ and $\Fm'\hskip-0.05em(\lt_i)$ are \emph{properly disjoint}. That is, $\lt_{i\hskip-0.07em} \wedge \Fm \rdc  \scp(\lt_i) \wedge \Fm'\hskip-0.05em(\lt_i)$  (see \cref{rdctn,rdctnTrm}). In this case, it is \emph{redundant} to check unsatisfiability of $\Fm'\hskip-0.05em(\lt_i)$ in order to decide unsatisfiability of $\Fm(\lt_i)$. This redundancy facilitates deciding satisfiability of $\FM$. Thus, \textit{X3SAT} leads to proving $\PP = \NP$.

\section[Decision Procedures]{Decision Procedures for (Quantified) Propositional Logic}
This chapter addresses the  reduction of $\FM$ to $\scp'\hskip-0.12em \wedge \Fm'\hskip-0.09em$. \cref{s:Unsat}  tackles unsatisfiability of $\FM$ and  \cref{s:Sat} tackles  satisfiability of $\Fm'\hskip-0.09em$.  \cref{s:Assg} addresses construction of  a satisfying assignment for $\Fm'\hskip-0.09em$. \cref{s:QxSAT} tackles \textit{TQBF}  via  ``prime satisfying assignments''.

\subsection[Unsatisfiability]{Unsatisfiability: Interruption of  Scan}\label{s:Unsat}
This section shows that inconsistency of a simple formula reduced from a formula is sufficient for the unsatisfiability of the formula. That is, $\unsat \FM$ if $\FM \prv \scp'\hskip-0.12em \wedge \Fm'\hskip-0.12em$ and  $\scp'\hskip-0.12em \prv x_{i\hskip-0.05em} \wedge \nx{i\hskip-0.05em}$. Note that it is trivial to check \emph{inconsistency}. Thus, it is easy to decide \emph{unsatisfiability}. See \cref{minterm,incnstnt,formula}.

\begin{definition}[Special formula]
$\Fm$  denotes a special formula if  $\{x_i, \nx{i}\} \subseteq C_{k\hskip-0.11em}$ for some $C_{k\hskip-0.05em}$.
\end{definition}

\begin{lemma}[Converting a special formula]\label{GnFm}
$\nlt_{j\hskip-0.17em}$ the conjunct replaces $(\lt_{j\hskip-0.09em} \X  x_{i\hskip-0.07em} \X \nx i)$ the clause.
\end{lemma}
\begin{proof}
$(\lt_{j\hskip-0.09em} \X  x_{i\hskip-0.07em} \X \nx i)$ is true by \cref{Xdsn} iff $(\lt_{j\hskip-0.09em} \wedge \nx{i\hskip-0.07em} \wedge x_i) \vee (\nlt_{j\hskip-0.09em} \wedge x_{i\hskip-0.07em} \wedge x_i) \vee (\nlt_{j\hskip-0.09em} \wedge \nx{i\hskip-0.07em} \wedge \nx i)$ is satisfiable. Therefore, the clause $(\lt_{j\hskip-0.09em} \X  x_{i\hskip-0.07em} \X \nx i)$ is true iff the literal $\nlt_{j\hskip-0.12em}$ becomes a conjunct.
\end{proof}

\begin{definition}\label{FmCk}
$\Fm^{\lt_i\hskip-0.12em} = \hskip-0.09em\bigwedge_{k \in \mathfrak{C}} C_{k\hskip-0.13em}$ such that $\lt_{i\hskip-0.13em} \in C_{k\hskip-0.05em}$, which can be empty, thus  $\mathfrak{C}^{\lt_i\hskip-0.17em} \subseteq  \mathfrak{C}$.
\end{definition}

\begin{example}
Let $\FM  = (x_{2\hskip-0.05em} \X \nx1) \wedge (x_{1\hskip-0.12em} \X \nx{3\hskip-0.05em} \X x_4) \wedge (x_{1\hskip-0.12em} \X \nx{2\hskip-0.05em} \X x_2)$, i.e., $\scp = \emptyset$ and $\mathfrak{C} = \{1, 2, 3\}$. Then, $3 \in (\mathfrak{C}^{\nx2\hskip-0.11em} \cap \mathfrak{C}^{x_2})$, where $\mathfrak{C}^{\nx2\hskip-0.11em} = \{3\}$ and $\mathfrak{C}^{x_2\hskip-0.11em} = \{1, 3\}$, i.e., $C_{3\hskip-0.07em}$ is contained  in $\Fm^{\nx2\hskip-0.09em}$ and  $\Fm^{x_2\hskip-0.07em}$. Hence, $\FM$ is converted by replacing  $C_{3\hskip-0.09em}$ with  $\nx{1\hskip-0.09em}$. Thus, $\FM \gets \nx{1\hskip-0.09em} \wedge (x_{2\hskip-0.05em} \X \nx1) \wedge (x_{1\hskip-0.12em} \X \nx3 \X x_4)$. Let $\FM =  (x_{3\hskip-0.05em} \X \nx{4\hskip-0.05em} \X x_4) \wedge (\nx{3\hskip-0.05em} \X x_{2\hskip-0.05em} \X \nx2) \wedge (x_{2\hskip-0.05em} \X \nx1)$. Then, $\FM \gets \nx{3\hskip-0.05em} \wedge x_{3\hskip-0.05em} \wedge (x_{2\hskip-0.05em} \X \nx1)$. Thus, $\unsat \FM$.
\end{example}

\begin{lemma}[Collapse of a formula]\label{Cllps}
$\lt_{i\hskip-0.07em} \wedge \Fm^{\lt_i\hskip-0.12em} \wedge  \Fm^{\nlt_i\hskip-0.12em} \prv \cllps(\lt_i)$, that is, $\Fm^{\lt_i\hskip-0.19em}$ collapses and becomes empty, in which $\cllps(\lt_i) = \hskip-0.09em\bigwedge_{k \in \mathfrak{C}^{\scriptstyle \lt_i\hskip-0.17em}} \scp_k(\lt_i) \wedge \bigwedge_{k \in \mathfrak{C}^{\scriptstyle \nlt_i\hskip-0.17em}} C_k(\neg \nlt_i)$ such that $C_k(\neg \nlt_i)$ is a unit clause.
\end{lemma}
\begin{proof}
Follows from \cref{cllps,shrnk}. Note that any unit clause becomes a conjunct.
\end{proof}

\begin{lemma}[Shrinkage of a formula]\label{Shrnk}
$\lt_{i\hskip-0.07em} \wedge \Fm^{\nlt_i\hskip-0.12em}  \prv \shrnk(\neg \nlt_i)$, and $\shrnk(\neg \nlt_i) = \hskip-0.09em\bigwedge_{k \in \mathfrak{C}^{\scriptstyle \nlt_i\hskip-0.17em}} C_k(\neg \nlt_i)$.
\end{lemma}
\begin{proof}
Follows from \cref{shrnk} such that $C_k(\neg \nlt_i)$ contains at least \emph{two} literals.
\end{proof}

\begin{remark*}
$\cllps(\lt_i)$ is to be consistent, while $\shrnk(\neg \nlt_i)$ can be empty since $|C_k| \geqslant 2$ by \cref{clause}.
\end{remark*}

\begin{lemma}[Reduction of a formula]\label{reduction}
$\lt_{i\hskip-0.07em} \wedge \Fm^{\lt_i\hskip-0.17em} \wedge  \Fm^{\nlt_i\hskip-0.12em}  \prv \rdctn(\lt_i)$, and $\rdctn(\lt_i) =  \cllps(\lt_i) \wedge \shrnk(\neg \nlt_i)$.
\end{lemma}
\begin{proof}
Follows from  \cref{Cllps,Shrnk}. Note that $\mathfrak{C}^{\lt_i\hskip-0.13em} \cap \mathfrak{C}^{\nlt_i\hskip-0.13em} = \emptyset$ due to  \cref{GnFm}.
\end{proof}

\begin{example}
Let  $\Fm = (x_{1\hskip-0.11em} \X \nx3) \wedge (x_{1\hskip-0.11em} \X \nx{2\hskip-0.05em} \X x_3) \wedge (x_{2\hskip-0.05em} \X \nx3)$ and $\scp = \emptyset$. Then, $\Li \cap \Li'\hskip-0.15em = \emptyset$ by \cref{LL'}. Thus,  \cref{ChkIncmp} is applicable. Hence, $\nx{i\hskip-0.07em}  \wedge \Fm \rdc  \scp(\nx i) \wedge \Fm'\hskip-0.05em(\nx i)$  for all $i \in \Li'\hskip-0.15em$ (see \cref{rdctn}, the second case, and \cref{rdctnTrm}). Next, consider $\Fm(x_1)$. Because $x_{1\hskip-0.15em} \in (C_{1\hskip-0.11em} \cap C_2)$,  $\Fm^{x_1\hskip-0.17em} =  C_{1\hskip-0.09em} \wedge C_{2\hskip-0.07em}$ by \cref{FmCk}. Then, $x_{1\hskip-0.11em} \wedge \Fm^{x_1\hskip-0.19em} \prv \cllps(x_1)$, and $x_{1\hskip-0.15em} \prv x_{3\hskip-0.03em} \wedge x_{2\hskip-0.02em} \wedge \nx{3\hskip-0.03em}$. Hence, $\unsat \Fm(x_1)$, thus $x_{1\hskip-0.17em}$ is \emph{incompatible} ($\nx{1\hskip-0.17em}$ is \emph{necessary}), i.e., $\neg x_{1\hskip-0.12em} \prv \nx{1\hskip-0.12em}$ and $\scp \gets \scp \wedge \nx{1\hskip-0.17em}$ by  \cref{NonTrvIncmp}. Note that $\nx{3\hskip-0.11em} \vee x_{3\hskip-0.09em} \imp \nx{1\hskip-0.05em}$. Consider $\Fm(x_3)$. $\Fm^{x_3\hskip-0.15em} = C_{2\hskip-0.03em}$, and $\Fm^{\nx3\hskip-0.15em} =   C_{1\hskip-0.07em} \wedge C_{3\hskip-0.03em}$. Then, $x_{3\hskip-0.03em} \wedge \Fm^{x_3\hskip-0.19em} \wedge \Fm^{\nx3\hskip-0.15em} \prv \cllps(x_3)$ by \cref{Cllps}, and $x_{3\hskip-0.03em} \wedge \Fm^{\nx3\hskip-0.15em} \prv \shrnk(\neg \nx3)$ by \cref{Shrnk}.  As a result, $\shrnk(\neg \nx3)$ is empty and $\cllps(x_3) \prv \linebreak \nx{1\hskip-0.11em} \wedge x_{2\hskip-0.05em} \wedge  C_1(\neg \nx3) \wedge C_3(\neg \nx3)$, i.e., $\cllps(x_3) \prv \nx{1\hskip-0.09em} \wedge x_{2\hskip-0.05em} \wedge  x_{1\hskip-0.11em} \wedge x_{2\hskip-0.03em}$. Hence, $\unsat \Fm(x_3)$, and $\scp \gets \scp \wedge \nx{3\hskip-0.03em}$.  Consider $\Fm(x_2)$. Then, $\cllps(x_2) = x_{3\hskip-0.03em}$, $\shrnk(\neg \nx2) = (x_{1\hskip-0.12em} \X  x_3)$, and $\rdctn(x_2) = \cllps(x_2) \wedge \shrnk(\neg \nx2)$. That is, $\rdctn(x_2) \hskip-0.05em= \nx{1\hskip-0.09em} \wedge  x_{3\hskip-0.03em}$. Then, $x_{2\hskip-0.03em} \wedge \Fm\hskip-0.03em \prv \rdctn(x_2) \wedge (x_{1\hskip-0.12em} \X \nx3)$. Hence, $\unsat \Fm(x_2)$, and $\scp \gets \scp \wedge \nx{2\hskip-0.03em}$.  Therefore,  $\scp = \nx{1\hskip-0.09em} \wedge \nx{3\hskip-0.05em} \wedge \nx{2\hskip-0.03em}$, and $\FM \gets \scp \wedge \Fm$. As a result, $\nx{1\hskip-0.15em}$ leads to reductions  (\cref{rdctn}, the first case), i.e., $\nx{1\hskip-0.07em} \wedge \Fm^{\nx1\hskip-0.15em} \wedge \Fm^{x_1\hskip-0.19em} \prv  \cllps(\nx1) \wedge \shrnk(\neg x_1)$ by  \cref{reduction}, where $\Fm^{\nx1\hskip-0.19em}$ is empty, $\cllps(\nx1) = C_1(\neg x_1) = \nx{3\hskip-0.03em}$, and $\shrnk(\neg x_1) =  C_2(\neg x_1) = (\nx{2\hskip-0.05em} \X x_3)$. Hence, $\FM \gets \scp \wedge \shrnk(\neg x_1) \wedge (x_{2\hskip-0.05em} \X \nx3)$. Finally, $\nx{3\hskip-0.05em}$ leads to reductions, i.e.,  $\nx{3\hskip-0.03em} \wedge \shrnk(\neg x_1) \wedge (x_{2\hskip-0.03em} \X \nx3) \prv \nx{2\hskip-0.03em} \wedge \nx{2\hskip-0.03em}$. Then, $\FM \hskip-0.05em\gets \nx{1\hskip-0.09em} \wedge \rdctn(\nx1) \wedge \nx{3\hskip-0.05em} \wedge \rdctn(\nx3)$. Thus, $\Fm$ reduces to the \emph{unique} satisfying assignment, viz., $\Fm \rdc   \nx{1\hskip-0.09em} \wedge \nx{3\hskip-0.05em} \wedge \nx{2\hskip-0.03em}$, i.e., $x_{1\hskip-0.13em} = 0, x_{2\hskip-0.09em} = 0, x_{3\hskip-0.09em} = 0$.
\end{example}

The algorithm \texttt{Reduce}~$\!(\Fm, \lt_j)$, specified below, constructs the reduction $\rdctn(\lt_j)$. It is due to the collapse $\cllps(\lt_j)$ (see Lines 1-8, or L1-8), or due to the shrinkage $\shrnk(\neg \nlt_j)$ (L9-16).

\begin{algorithm}\caption{\texttt{Reduce}~$\!(\Fm, \lt_j)$ \Cmmnt{Construction of the reduction by \cref{reduction}, $\lt_{j\hskip-0.07em} \wedge \Fm^{\lt_j\hskip-0.17em} \wedge  \Fm^{\nlt_j\hskip-0.12em}  \prv \rdctn(\lt_j)$}}
\begin{algorithmic}[1]
\ForAll {$k \in \mathfrak{C}^{\lt_j\hskip-0.15em}$} \Cmmnt{$|\mathfrak{C}^{\lt_j}| \leqslant m$ by \cref{FmCk}}
    \ForAll {$\lt_{i \neq j \hskip-0.12em} \in C_{k\hskip-0.09em}$} \Cmmnt{$|C_k| \leqslant 3$ for \textit{X3SAT} and $|C_k| \leqslant n$ for \textit{XSAT} by \cref{clause}}
        \State $\scp_k(\lt_j) \gets \scp_k(\lt_j) \wedge \nlt_{i\hskip-0.17em}$ \Cmmnt{\hskip-0.09em$\scp_k(\lt_j) \hskip-0.03em= \nlt_{i\hskip-0.07em} \wedge \cdots \wedge \nlt_{u\hskip-0.15em}$ for $C_{k\hskip-0.12em} =\hskip-0.03em (\lt_{i\hskip-0.07em} \X \lt_{j\hskip-0.09em} \X \cdots \X \lt_u)\hskip-0.05em$ (see  \cref{cllps})}
    \EndFor \Cmmnt{$\lt_{j\hskip-0.07em} \wedge C_{k\hskip-0.09em} \prv \scp_k(\lt_j)$ by  \cref{cllps}\,---\,the clause $C_{k\hskip-0.12em}$ collapses to the minterm $\scp_k(\lt_j)$}
    \State $\cllps(\lt_j) \gets \cllps(\lt_j) \wedge \scp_k(\lt_j)$ \Cmmnt{Construction of the collapse by \cref{Cllps} due to $\bigwedge_{k \in \mathfrak{C}^{\scriptstyle \lt_j\hskip-0.17em}} \scp_k(\lt_j)$}
    \State \textbf{if} $\cllps(\lt_j) \prv x_{i\hskip-0.09em} \wedge \nx{i\hskip-0.09em}$ for some $i$ \textbf{then} \Return $\cllps(\lt_j)$ is inconsistent \Cmmnt{See also \cref{incnstnt}}
    \State Remove $C_{k\hskip-0.09em}$ from $\Fm^{\lt_j\hskip-0.12em}$ \Cmmnt{The clause $C_{k\hskip-0.12em}$ collapsed in $\Fm^{\lt_j\hskip-0.17em}$  becomes empty (see  \cref{cllps})}
\EndFor \Cmmnt{$\lt_{j\hskip-0.07em} \wedge \Fm^{\lt_j\hskip-0.12em}  \prv \cllps(\lt_j)$ by \cref{Cllps}---\,the formula $\Fm^{\lt_j\hskip-0.19em}$ collapsed becomes empty due to L7}
\ForAll {$k \in \mathfrak{C}^{\nlt_j\hskip-0.15em}$} \Cmmnt{$|\mathfrak{C}^{\nlt_j}| \leqslant m$ by \cref{FmCk}}
    \State Remove $\nlt_{j\hskip-0.12em}$ from $C_{k\hskip-0.19em}$ \Cmmnt{Construction of the shrinkage (\cref{Shrnk}), $\shrnk(\neg \nlt_j) = \hskip-0.09em\bigwedge_{k \in \mathfrak{C}^{\scriptstyle \nlt_j\hskip-0.17em}} C_k(\neg \nlt_j)$}
    \If {$C_{k\hskip-0.09em} = (\lt_i)$ and $\lt_{i\hskip-0.09em} \notin  \cllps(\lt_j)$} \Cmmnt{$\lt_{j\hskip-0.09em} \wedge C_{k\hskip-0.09em}  \prv C_k(\neg \nlt_j)$ by  \cref{shrnk}, and $C_k(\neg \nlt_j) = (\lt_i)$}
    	\State $\cllps(\lt_j) \gets \cllps(\lt_j) \wedge \lt_{i\hskip-0.12em}$ \Cmmnt{Construction of the collapse by \cref{Cllps} due to $\bigwedge_{k \in \mathfrak{C}^{\scriptstyle \nlt_j\hskip-0.17em}} C_k(\neg \nlt_j)$}
	    \State \textbf{if} $\cllps(\lt_j) \prv x_{i\hskip-0.07em} \wedge \nx{i\hskip-0.09em}$ \textbf{then} \Return $\cllps(\lt_j)$ is inconsistent
	    \State Remove $C_{k\hskip-0.12em}$ from $\Fm^{\nlt_j\hskip-0.12em}$ \Cmmnt{$C_{k\hskip-0.12em}$  becomes empty by  \cref{shrnk} ($|C_k| \geqslant 2$ by  \cref{clause})}
   \EndIf
\EndFor \Cmmnt{$\lt_{j\hskip-0.09em} \wedge \Fm^{\nlt_j\hskip-0.12em}  \prv \shrnk(\neg \nlt_j)$ by \cref{Shrnk}. $\Fm^{\nlt_j\hskip-0.12em}$  becomes empty, or $|C_k| \geqslant 2$ for each $C_{k\hskip-0.12em}$ in $\Fm^{\nlt_j\hskip-0.12em}$}
\State \Return $\cllps(\lt_j) \wedge \Fm'$ \Cmmnt{$\lt_{j\hskip-0.07em} \wedge \Fm \prv \cllps(\lt_j) \wedge \Fm'\hskip-0.09em$, $\Fm \supseteq \Fm'\hskip-0.09em$, i.e., $\mathfrak{C} \supseteq \mathfrak{C}'\hskip-0.17em$ by L7/14 \& $\forall k' \mspace{1mu} \exists k \, C_{k\hskip-0.09em} \supseteq C_{k'\hskip-0.17em}$ by L10. $\Fm'\hskip-0.15em  = \shrnk(\neg \nlt_j) \wedge \bar{\Fm}$. $\shrnk(\neg \nlt_j) = \Fm^{\nlt_j\hskip-0.12em}$.  $\bar{\Fm} =  \hskip-0.09em\bigwedge_{k \in \bar{\mathfrak{C}}} C_{k\hskip-0.05em}$,  $\bar{\mathfrak{C}} = \mathfrak{C} - (\mathfrak{C}^{\lt_j\hskip-0.15em} \cup \mathfrak{C}^{\nlt_j})$. $\mathfrak{C}^{\lt_j\hskip-0.15em} \cap \mathfrak{C}^{\nlt_j\hskip-0.15em} = \emptyset$ by \cref{GnFm}}
\end{algorithmic}
\label{alg:Reduce}
\end{algorithm}

\texttt{Scope}~$\!(\lt_j, \Fm)$  decides \emph{nontrivial} incompatibility  (L4,7). See also \cref{noScope}. \texttt{Scope}~$\!(\lt_i, \Fm)$ constructs the scope $\scp(\lt_i)$ (L9,12), and  the beyond the scope $\Fm'\hskip-0.05em(\lt_i)$. See also \cref{Scope}.

\begin{algorithm}\caption{\texttt{Scope}~$\!(\lt_j, \Fm)$ \Cmmnt{Inconsistency of $\scp(\lt_j)$ (\cref{noScope})-Consistency of  $\scp(\lt_i)$ (\cref{Scope})}}
\begin{algorithmic}[1]
\State $\scp(\lt_j) \gets \lt_j$; $\Fm'\hskip-0.05em(\lt_j) \gets \Fm$ \Cmmnt{$\Fm(\lt_j) = \lt_{j\hskip-0.07em} \wedge \Fm$ initially by \cref{ChkIncmp}. $\Fm$ is nonempty by \texttt{Scan} \hyperref[alg:Scan]{L10}}
\ForAll {$\lt_{j\hskip-0.09em} \in \scp(\lt_j)$} \Cmmnt{Initiation of the reductions over $\Fm'\hskip-0.05em(\lt_j)$. $|\scp(\lt_j)| \leqslant n$ by  \cref{literal}}
\State \texttt{Reduce}~$\!\big(\Fm'\hskip-0.05em(\lt_j), \lt_j\big)$ \Cmmnt{Returns $\cllps(\lt_j)$ and $\Fm'\hskip-0.09em$. See   \cref{rdctn}, the second case}
       \If {$\cllps(\lt_j)$ is inconsistent} \Return NULL  \Cmmnt{\cref{noScope}}
       \ElsIf {$\cllps(\lt_j)$ is nonempty} \Cmmnt{It is empty if $\Fm^{\lt_j\hskip-0.12em}$ is empty and $|C_k| > 2$ for all $C_{k\hskip-0.12em}$ in $\Fm^{\nlt_j\hskip-0.12em}$}
          \State $\scp(\lt_j) \gets \scp(\lt_j) \wedge \cllps(\lt_j)$ \Cmmnt{Construction of the scope $\scp(\lt_j)$. It is to be consistent}
          \State \textbf{if} $\scp(\lt_j)$ is inconsistent \textbf{then} \Return NULL \Cmmnt{\cref{noScope}. $\lt_{j\hskip-0.07em} \wedge \Fm \prv \scp(\lt_j) \wedge \Fm'\hskip-0.05em(\lt_j)$ and $\unsat \Fm(\lt_j)$, thus $\lt_{j\hskip-0.12em}$ is incompatible for $\Fm$ and $\nlt_{j\hskip-0.12em}$ is necessary for $\FM$. $j \in \ell$ by \cref{NonTrvIncmp,NonTrvNss}}
       \EndIf
       \State \textbf{if} $\Fm'\hskip-0.12em$ is  empty \textbf{then} \Return $\scp(\lt_i)$ \Cmmnt{$\lt_{i\hskip-0.07em} \wedge \Fm \rdc \scp(\lt_i)$ (cf. \!L12). $i \in \Li'\hskip-0.12em$ by \cref{CmpL}, $i = j$}
       \State $\Fm'\hskip-0.05em(\lt_j) \gets \Fm'\hskip-0.12em$ \Cmmnt{$\Fm'\hskip-0.05em(\lt_j)$ is updated. It also involves the unreduced clauses, denoted by $\bar{\Fm}$}
\EndFor \Cmmnt{\cref{Indp,Scope}. \cref{subset}. Termination of the reductions over $\Fm'\hskip-0.05em(\lt_i)$. $i \in \Li'\hskip-0.12em$}
\State \Return $\scp(\lt_i) \wedge \Fm'\hskip-0.05em(\lt_i)$ \Cmmnt{$\lt_{i\hskip-0.07em} \wedge \Fm \rdc  \scp(\lt_i) \wedge \Fm'\hskip-0.05em(\lt_i)$. $\Fm \supseteq  \Fm'\hskip-0.05em(\lt_i)$. $\scp(\lt_i)$  \& $\Fm'\hskip-0.05em(\lt_i)$ are properly disjoint}
\end{algorithmic}
\label{alg:Scope}
\end{algorithm}

\begin{lemma}[Nontrivial incompatibility before the scan termination]\label{noScope}
If $\scp(\lt_j)$ is inconsistent, then $\unsat \Fm(\lt_j)$, thus $\scp  \gets \scp \wedge \nlt_{j\hskip-0.05em}$, that is, $\lt_{j\hskip-0.15em}$ is incompatible for $\Fm$,  thus  $\nlt_{j\hskip-0.15em}$ is necessary for $\FM$.
\end{lemma}
\begin{proof}
$\Fm(\lt_j) = \lt_{j\hskip-0.07em} \wedge \Fm$ by \cref{ChkIncmp}. Hence,  $\lt_{j\hskip-0.07em} \wedge \Fm \prv \scp(\lt_j) \wedge \Fm'\hskip-0.05em(\lt_j)$ (see \texttt{Scope} L6,10). As a result, if $\scp(\lt_j)$ is inconsistent, then $\unsat \Fm(\lt_j)$ (L4,7). Thus, $\scp  \gets  \scp \wedge \nlt_{j\hskip-0.12em}$ by \cref{NonTrvIncmp}.
\end{proof}

\texttt{Scan}~$\!(\FM)$ is specified below. As $\nlt_{j\hskip-0.12em} \in \scp$, $\lt_{j\hskip-0.12em}$  is incompatible \emph{trivially} by \cref{TrvIncmp}, and $\Fm$ is reduced by $\nlt_{j\hskip-0.12em}$ (L2-12).  As $\unsat \Fm(\lt_j)$, $\lt_{j\hskip-0.12em}$  is incompatible \emph{nontrivially} by \cref{NonTrvIncmp}   (L15-22).

\newpage

\begin{algorithm} \caption{\texttt{Scan}~$\!(\FM)$ \Cmmnt{$\FM = \scp \wedge \Fm$ initially. \texttt{Scan}~$\!(\FM_s)$  runs over $s = 0, 1, \ldots, n$ (L2-22) unless $\unsat \FM$}}
\begin{algorithmic}[1]
\Repeat \Cmmnt{\texttt{Scan}~$\!(\FM_{s+1})$ runs whenever the empty $\scp_{s\hskip-0.12em}$ (see L13) becomes a nonempty $\scp_{s+1\hskip-0.15em}$ (L18)}
\ForAll {$\nlt_{j\hskip-0.09em} \in \scp$} \Cmmnt{$\nlt_{j\hskip-0.12em}$ initiates a new cycle of reductions over $\Fm$. $\nlt_{j\hskip-0.12em}$ is in $\scp_{s+1\hskip-0.09em}$ by L18}
    \State \texttt{Reduce}~$\!(\Fm, \nlt_j)$ \Cmmnt{Returns $\cllps(\nlt_j)$ and $\Fm'\hskip-0.09em$. See   \cref{rdctn}, the first case}
	\If {$\cllps(\nlt_j)$ is inconsistent}  \Return UNSAT \Cmmnt{$\FM \prv \cllps(\nlt_j) \wedge \Fm'\hskip-0.09em$. See also \cref{incnstnt}}
	\ElsIf {$\cllps(\nlt_j)$ is nonempty}
	   \State $\scp \gets \scp \wedge \cllps(\nlt_j)$ \Cmmnt{Let $\nlt_{i\hskip-0.12em} \in \cllps(\nlt_j)$. $\nlt_{j\hskip-0.09em} \prv \nlt_{i\hskip-0.15em}$ ($\nlt_{i\hskip-0.12em}$ is \textbf{trivially} necessary by \cref{TrvNss})}
	   \State $\scp'\hskip-0.09em \gets \scp'\hskip-0.12em \wedge \scp$ \Cmmnt{$\scp'\hskip-0.09em$ denotes the conjuncts that have already reduced the formula $\Fm$}
	   \State \textbf{if} $\scp'\hskip-0.12em$ is inconsistent \textbf{then} \Return UNSAT \Cmmnt{$\FM \prv \scp'\hskip-0.12em \wedge \Fm'\hskip-0.09em$. See also \cref{incnstnt}}
	\EndIf
	\State \textbf{if} $\Fm'\hskip-0.12em$ is empty \textbf{then}  \Return  $\scp'\hskip-0.12em$ \Cmmnt{Termination, $\FM \rdc \scp'\hskip-0.15em$ (unique satisfying assignment)}
	\State $\Fm \gets \Fm'\hskip-0.09em$ \Cmmnt{$\Fm'\hskip-0.15em = \shrnk(\neg \lt_j) \wedge \bar{\Fm}$.  $\bar{\Fm} =  \hskip-0.09em\bigwedge_{k \in \bar{\mathfrak{C}}} C_{k\hskip-0.05em}$,  $\bar{\mathfrak{C}} = \mathfrak{C} - (\mathfrak{C}^{\nlt_j\hskip-0.15em} \cup \mathfrak{C}^{\lt_j})$. $\mathfrak{C}^{\nlt_j\hskip-0.15em} \cap \mathfrak{C}^{\lt_j\hskip-0.15em} = \emptyset$ by \cref{GnFm}}
\EndFor \Cmmnt{This cycle of the reductions over the \emph{current} $\Fm$ terminates, i.e., $\scp \wedge \Fm \rdc \scp'\hskip-0.12em \wedge \Fm'\hskip-0.12em$}
\State $\scp \gets \emptyset$ \Cmmnt{$\scp$ is reset and $\Fm'\hskip-0.12em$ becomes the initial formula. $\scp_0, \ldots, \scp_{n\hskip-0.09em}$ become properly disjoint}
\State $\Li'\hskip-0.15em = \mathfrak{\Li} - \Li$  \hskip-0.09em\Cmmnt{$\Li = \{ j \mid \lt_{j\hskip-0.12em} \in \scp'\}$ by \cref{LL'}.  $\scp'\hskip-0.09em$ and $\Fm'\hskip-0.09em$ are properly disjoint as $\FM \rdc \scp'\hskip-0.12em \wedge \Fm'\hskip-0.09em$}
\ForAll {$i \in \Li'\hskip-0.12em$} \Cmmnt{A new cycle of incompatibility checking over $\Fm$ starts off by \cref{ChkIncmp}}
    \ForAll {$\lt_{i\hskip-0.12em} \in \{x_i, \nx i\}$}
    	\If {\texttt{Scope}~$\!(\lt_i, \Fm)$ is NULL} \Cmmnt{$\lt_{i\hskip-0.12em}$ is \textbf{nontrivially} incompatible by \cref{NonTrvIncmp}}
    	  \State $\scp  \gets \scp \wedge \nlt_{i\hskip-0.15em}$ \Cmmnt{$\neg \lt_{i\hskip-0.11em} \prv \nlt_{i\hskip-0.13em}$ ($\nlt_{i\hskip-0.15em}$ is \textbf{nontrivially} necessary by \cref{NonTrvNss,NonTrvIncmp})}
	      \State \textbf{if} $\scp$ is inconsistent \textbf{then} \textbf{return} UNSAT  \Cmmnt{$\FM \prv \scp \wedge \Fm'\hskip-0.09em$. See also \cref{incnstnt}}
   	   \EndIf
    \EndFor
\EndFor \Cmmnt{This cycle of the incompatibility checking over the \emph{current} $\Fm$ terminates}
\Until{$\scp = \emptyset$} \!\Cmmnt{Reductions (L2-12) and  incompatibility checking (L15-22) are mutually exclusive. Thus, they can be in an arbitrary order. Also, \texttt{Scan}~$\!(\FM_0),\ldots,$ \texttt{Scan}~$\!(\FM_n)$ are mutually exclusive}
\State \Return $\FM'\hskip-0.12em$ \Cmmnt{Termination, $\FM \rdc \scp'\hskip-0.12em \wedge \Fm'\hskip-0.12em$ (cf. \!L10). Construction of a satisfying assignment over $\Fm'\hskip-0.12em$}
\end{algorithmic}
\label{alg:Scan}
\end{algorithm}

This section showed that $\FM$ is unsatisfiable if \texttt{Scan}~$\!(\FM)$ is interrupted (see L4,8,19).

\subsection[Satisfiability]{Satisfiability: Termination of  Scan}\label{s:Sat}
This section shows that $\FM$ is satisfiable if \texttt{Scan}~$\!(\FM)$ terminates due to L24.  The proof  is to show reducibility of $\Fm'\hskip-0.12em$ the formula to a minterm $\hat{\scp}$  consistent, i.e., $\Fm'\hskip-0.12em \rdc \hat{\scp}$. Let $\FM \gets \FM'\hskip-0.09em$.

\begin{lemma}[Scope]\label{Scope}
Termination of the reductions due to $\lt_{i\hskip-0.11em}$ over $\Fm$ results in the scope $\scp(\lt_i)$, $\scp(\lt_i) = \hskip-0.09em\bigwedge\hskip-0.09em \big(\lt_{i\hskip-0.07em} \wedge \cllps(\lt_i)\big)$, viz., $\lt_{i\hskip-0.07em} \wedge \Fm \rdc  \scp(\lt_i) \wedge \Fm'\hskip-0.05em(\lt_i)$ for any $i \in \Li'\hskip-0.11em$ and each $\lt_{i\hskip-0.09em} \in X_{i\hskip-0.05em}$.
\end{lemma}
\begin{proof}
Follows from \texttt{Scope} \hyperref[alg:Scope]{L6,12}. See also \cref{ChkIncmp,rdctnTrm}.
\end{proof}

\begin{remark}\label{subset}
\cref{Scope} also entails the reduction of  $\Fm$  to $\Fm'\hskip-0.05em(\lt_i)$, viz., $\Fm \supseteq \Fm'\hskip-0.05em(\lt_i)$. That is,  $\mathfrak{C} \supseteq \mathfrak{C}'\hskip-0.17em$ and $\forall k' \mspace{1mu} \exists k \, C_{k\hskip-0.09em} \supseteq C_{k'\hskip-0.09em}$, where $\mathfrak{C}'\hskip-0.15em$ denotes the clauses $C_{k\hskip-0.12em}$ in $\Fm'\hskip-0.05em(\lt_i)$. See also \texttt{Reduce}~$\!(\Fm, \lt_i)$.
\end{remark}

Let $\Li(\lt_i) \hskip-0.05em=\hskip-0.01em \Li(\scp(\lt_i))$ and $\Li'\hskip-0.05em(\lt_i) \hskip-0.05em=\hskip-0.01em  \Li(\Fm'\hskip-0.05em(\lt_i))$, e.g., $\Li(x_3) \hskip-0.05em=\hskip-0.01em \{3, 4, 5\}$ and $\Li'\hskip-0.05em(x_3) \hskip-0.05em=\hskip-0.01em  \{1, 2, 6,  7\}$ due to  $\scp(x_3) = x_{3\hskip-0.05em} \wedge \nx{4\hskip-0.06em} \wedge x_{5\hskip-0.05em}$ and $\Fm'\hskip-0.05em(x_3) =  (x_{1\hskip-0.12em} \X \nx{2\hskip-0.05em} \X \nx6) \wedge  (x_{6\hskip-0.05em} \X \nx7)$. Then, $\Li(x_3) \cap \Li'\hskip-0.05em(x_3) = \emptyset$.

\begin{lemma}\label{Indp}
$\Li(\lt_i) \cap \Li'\hskip-0.05em(\lt_i) = \emptyset$, that is, $\scp(\lt_i)$  and $\Fm'\hskip-0.05em(\lt_i)$  are \emph{properly} disjoint.
\end{lemma}
\begin{proof}
Follows from \cref{Scope}. See  \cref{PrpDsj}. Let $\Fm \hskip-0.09em\coloneqq \Fm'\hskip-0.05em(\lt_i)$. Let $u \in \big(\Li(\lt_i) \cap \Li'\hskip-0.05em(\lt_i)\big)$. Then, $\lt_{u\hskip-0.13em} \in \scp(\lt_i)$, which is  a \emph{conjunct} by \cref{minterm}, thus $\lt_{u\hskip-0.05em} \wedge \Fm^{\lt_u} \hskip-0.15em \prv \cllps(\lt_u)$ by  \cref{Cllps} (see also \texttt{Reduce} L1-8) in order to construct the scope  $\scp(\lt_i)$ (see \texttt{Scope} L6). Hence, $\Fm^{\lt_u}\hskip-0.15em$ becomes empty (see \texttt{Reduce} L7). Thus, $\lt_{u\hskip-0.13em} \notin C_{k\hskip-0.12em}$ for any clause $C_{k\hskip-0.12em}$ in $\Fm'\hskip-0.05em(\lt_i)$. Also, $\lt_{u\hskip-0.05em} \wedge \Fm^{\nlt_u} \hskip-0.15em \prv \tilde{\Fm}(\neg \nlt_u)$  due to  \cref{Shrnk} (\texttt{Reduce} L9-16). Thus,  $\nlt_{u\hskip-0.11em} \notin C_{k\hskip-0.12em}$ for any $C_{k\hskip-0.12em}$ in $\Fm'\hskip-0.05em(\lt_i)$.  Therefore, $u \notin \Li'\hskip-0.05em(\lt_i)$.
\end{proof}

\begin{definition}[Nontrivial necessity]\label{NonTrvNss}
$\ell = \{j \in \Li \mid \scp(\lt_j) \text{ is inconsistent for some } \lt_{j\hskip-0.09em} \in  X_j \}$.
\end{definition}

\begin{definition}[Trivial necessity]\label{TrvNss}
$\bar{\ell} =  \{ j'\hskip-0.09em \in \Li \mid j \in \ell \text{ but } j'\hskip-0.09em \notin \ell \text{ and }  \nlt_{j\hskip-0.09em} \prv \lt_{j'} \}$.
\end{definition}

\begin{definition}[Compatibility]\label{CmpL}
$\Li'\hskip-0.15em =  \{i \in \mathfrak{\Li} \mid \scp(\lt_i) \text{ is consistent for each } \lt_{i\hskip-0.09em} \in  X_i \}$.
\end{definition}

\begin{note}\label{NonTrv}
$\Li = \{ j \mid \lt_{j\hskip-0.12em} \in \scp\}$ and  $\Li'\hskip-0.15em = \{ i \mid \lt_{i\hskip-0.09em} \in C_{k\hskip-0.12em} \text{ for some } C_{k\hskip-0.09em} \text { in } \Fm\}$ by \cref{LL'}. If $\scp(\lt_j)$ is inconsistent, then $\unsat \Fm(\lt_j)$ and $\scp \gets \scp \wedge \nlt_{j\hskip-0.12em}$ by \cref{noScope}. Thus, $j \in \ell$ by \cref{LL',NonTrvNss}.  Also, if $\nlt_{j\hskip-0.11em} \prv \lt_{j'\hskip-0.07em}$, then $\scp \gets \scp \wedge \lt_{j'\hskip-0.12em}$. Thus, $j' \hskip-0.11em \in \bar{\ell}$ by \cref{TrvNss}. Then, $\Li$ (and $\scp$) is constructed either via \texttt{Scan} L17-18  or  L2-6. On the other hand, if $\scp$ is nonempty initially (see \cref{formula}), then $v \in \bar{\ell}$ such that $\lt_{t\hskip-0.11em} \in \scp$ and $\lt_{t\hskip-0.11em} \prv \lt_{v\hskip-0.05em}$. See also \cref{TrvIncmp,NonTrvIncmp}.
\end{note}

\begin{lemma}\label{TRmnt} $\Li \cap \Li'\hskip-0.15em = \emptyset$, as well as $\ell \cap \bar{\ell}  = \emptyset$ and $\ell \cup \bar{\ell}  = \Li$, when the $\FM$ scan terminates.
\end{lemma}
\begin{proof}
Follows directly from \crefrange{NonTrvNss}{CmpL}. Recall that $\Li \cup \Li'\hskip-0.15em = \mathfrak{\Li}$  by \cref{LL'}.
\end{proof}

\begin{remark*}
\cref{TRmnt} entails that $\Li'\hskip-0.09em$ \emph{becomes} the complement  of $\Li$ when the scan \emph{terminates}.
\end{remark*}

If $\scp(\lt_j)$ is inconsistent, then $\unsat \Fm(\lt_j)$ and $\lt_{j\hskip-0.12em}$ is removed from $\Fm$,  \emph{before} the termination of \texttt{Scan} (see \cref{noScope}). Thus,   $\scp(\lt_i)$ is consistent and  $\Fm(\lt_i) = \scp(\lt_i) \wedge \Fm'\hskip-0.05em(\lt_i)$ for any $i \in \Li'\hskip-0.09em$ and $\lt_{i\hskip-0.09em} \in X_{i\hskip-0.05em}$, \emph{after} the termination (\cref{Scope}).  Then, whether or not $\unsat \Fm'\hskip-0.05em(\lt_i)$ is to be checked in order to decide \emph{nontrivial} incompatibility of $\lt_{i\hskip-0.05em}$, i.e., to decide if $\unsat \Fm(\lt_i)$ (see \cref{void}).

\begin{lemma}[Nontrivial incompatibility after the scan termination]\label{void}
$\unsat \Fm(\lt_i)$ iff $\unsat \Fm'\hskip-0.05em(\lt_i)$.
\end{lemma}
\begin{proof}
Follows directly from \cref{Scope}. See also \cref{Indp}.
\end{proof}

\begin{claim}[Incompatibility assumption]\label{assmp}
It is \emph{redundant} to check whether $\unsat \Fm'\hskip-0.05em(\lt_i)$ in order to decide incompatibility of $\lt_{i\hskip-0.05em}$, i.e., to decide if $\unsat \Fm(\lt_i)$. Thus, $\unsat \Fm(\lt_j)$ iff $\scp(\lt_j)$ is inconsistent.
\end{claim}

\begin{remark*}
Satisfiability of $\Fm$ by \cref{final} justifies \cref{assmp}, thus \cref{void} becomes \emph{void}. Recall that $\scp \wedge \Fm \rdc \scp'\hskip-0.12em \wedge \Fm'\hskip-0.12em$ (see \texttt{Scan} L24), and that $\Fm \gets \Fm'\hskip-0.09em$. That is, $\Fm'\hskip-0.12em$ is the current formula after the termination. Also, it is \emph{trivial} to check  inconsistency of $\scp(\lt_j)$ by \cref{incnstnt}.
\end{remark*}

\begin{definition}\label{ConScp}
$\scp(\lt_{i_1} |\, \lt_{i_0})$ denotes a conditional  scope due to $\lt_{i_1\hskip-0.27em}$ over $\Fm'\hskip-0.05em(\lt_{i_0})$,  constructed  via \texttt{\emph{Scope}}~$\!\big(\lt_{i_1\hskip-0.05em}, \Fm'\hskip-0.05em(\lt_{i_0})\big)$. Likewise,  $\scp(\lt_{i_k} |\, \lt_{i_{k-1}})$ is due to $\lt_{i_k\hskip-0.27em}$ over $\Fm'\hskip-0.05em(\lt_{i_{k-1}} |\, \lt_{i_{k-2}})$ for $k = 2, 3, \ldots, n$.
\end{definition}

\begin{lemma}[Recursive reductions]\label{prdctns}
$\Fm \rdc \scp(\lt_{i_0}) \wedge \Fm'\hskip-0.05em(\lt_{i_0}), \, \Fm'\hskip-0.05em(\lt_{i_0})  \rdc \scp(\lt_{i_1} |\, \lt_{i_0}) \wedge \Fm'\hskip-0.05em(\lt_{i_1} |\, \lt_{i_0}), \linebreak \ldots,  \Fm'\hskip-0.05em(\lt_{i_{n-1}} |\, \lt_{i_{n-2}}) \rdc  \scp(\lt_{i_n} |\, \lt_{i_{n-1}})$, in which $\scp(.)$ and  $\Fm'\hskip-0.05em(.)$ are properly disjoint, that is, $\Fm \rdc \scp(\lt_{i_0}) \wedge  \scp(\lt_{i_1} |\, \lt_{i_0}) \wedge \cdots \wedge  \scp(\lt_{i_n} |\, \lt_{i_{n-1}})$ and $\Fm \supseteq \Fm'\hskip-0.05em(\lt_{i_0}) \supseteq \Fm'\hskip-0.05em(\lt_{i_1} |\, \lt_{i_0})  \supseteq \cdots \supseteq \Fm'\hskip-0.05em(\lt_{i_{n-1}} |\, \lt_{i_{n-2}})$, for any $\lt_{i_k\hskip-0.23em} \in X_{i_k\hskip-0.05em}$, and  $i_{0\hskip-0.15em} \in \Li'\hskip-0.09em, \, i_{1\hskip-0.15em} \in \Li'\hskip-0.05em(\lt_{i_0}),  \ldots,  i_{k\hskip-0.11em} \in \Li'\hskip-0.05em(\lt_{i_{k-1}} |\, \lt_{i_{k-2}})$, where $\Li'\hskip-0.05em(\lt_{i_n} |\, \lt_{i_{n-1}}) = \emptyset$.
\end{lemma}
\begin{proof}
Follows from \cref{Scope,ConScp}. See  also \cref{Indp}, as well as \cref{subset}.  Firstly, $\lt_{i_0\hskip-0.11em} \wedge \Fm \rdc \scp(\lt_{i_0}) \wedge \Fm'\hskip-0.05em(\lt_{i_0})$ and $\lt_{i_1\hskip-0.19em} \wedge \Fm'\hskip-0.05em(\lt_{i_0})  \rdc \scp(\lt_{i_1} |\, \lt_{i_0}) \wedge \Fm'\hskip-0.05em(\lt_{i_1} |\, \lt_{i_0})$ via \texttt{Scope}~$\!(\lt_{i_0\hskip-0.05em}, \Fm)$ and \texttt{Scope}~$\!\big(\lt_{i_1\hskip-0.05em}, \Fm'\hskip-0.05em(\lt_{i_0})\big)$, respectively. Likewise,  $\Fm'\hskip-0.05em(\lt_{i_{k-1}} |\, \lt_{i_{k-2}}) \rdc \scp(\lt_{i_k} |\, \lt_{i_{k-1}}) \wedge \Fm'\hskip-0.05em(\lt_{i_k} |\, \lt_{i_{k-1}})$ for $k = 2, 3, \ldots, n-1$. Finally,  $\Fm'\hskip-0.05em(\lt_{i_{n-1}} |\, \lt_{i_{n-2}}) \rdc \scp(\lt_{i_n} |\, \lt_{i_{n-1}})$, that is, $\Fm'\hskip-0.05em(\lt_{i_n} |\, \lt_{i_{n-1}})$ is empty. Therefore, $\Fm \supseteq \Fm'\hskip-0.05em(\lt_{i_0}) \supseteq \Fm'\hskip-0.05em(\lt_{i_1} |\, \lt_{i_0})  \supseteq \cdots \supseteq \Fm'\hskip-0.05em(\lt_{i_{n-1}} |\, \lt_{i_{n-2}})$, in which $\Fm'\hskip-0.05em(\lt_{i_0}) \supseteq \Fm'\hskip-0.05em(\lt_{i_1} |\, \lt_{i_0})$ via  \texttt{Reduce}~$\!\big(\Fm'\hskip-0.05em(\lt_{i_0}), \lt_{i_1\hskip-0.05em}\big)$, and  $\Fm'\hskip-0.05em(\lt_{i_{k-1}} |\, \lt_{i_{k-2}}) \supseteq  \Fm'\hskip-0.05em(\lt_{i_k} |\, \lt_{i_{k-1}})$ via \texttt{Reduce}~$\!\big(\Fm'\hskip-0.05em(\lt_{i_{k-1}} |\, \lt_{i_{k-2}}), \lt_{i_k}\big)$.
\end{proof}

\begin{lemma}[Any conditional scope is a syntactic consequence of its scope]\label{prv}
For each $\lt_{i_k\hskip-0.23em} \in X_{i_k\hskip-0.13em}$, $\scp(\lt_{i_1\hskip-0.05em}) \prv \scp(\lt_{i_1} |\, \lt_{i_0})$ for all $i_{1\hskip-0.15em} \in \Li'\hskip-0.05em(\lt_{i_0})$, and $\scp(\lt_{i_k}) \prv \scp(\lt_{i_k} |\, \lt_{i_{k-1}})$ for all $i_{k\hskip-0.07em} \in \Li'\hskip-0.05em(\lt_{i_{k-1}} |\, \lt_{i_{k-2}})$.
\end{lemma}
\begin{proof}
Follows directly from \cref{prdctns}. $\Fm \supseteq \Fm'\hskip-0.05em(\lt_{i_0})$, $\Fm \rdc \scp(\lt_{i_1})$, and $\Fm'\hskip-0.05em(\lt_{i_0}) \rdc \scp(\lt_{i_1} |\, \lt_{i_0})$. Thus, $\scp(\lt_{i_1\hskip-0.05em}) \supseteq \scp(\lt_{i_1} |\, \lt_{i_0})$. Hence, $\scp(\lt_{i_1\hskip-0.05em}) \prv \scp(\lt_{i_1} |\, \lt_{i_0})$. Also, $\Fm \supseteq \Fm'\hskip-0.05em(\lt_{i_1} |\, \lt_{i_0})$,  $\Fm \rdc \scp(\lt_{i_2})$, and $\Fm'\hskip-0.05em(\lt_{i_1} |\, \lt_{i_0}) \rdc \scp(\lt_{i_2} |\, \lt_{i_1})$. Thus, $\scp(\lt_{i_2\hskip-0.05em}) \supseteq \scp(\lt_{i_2} |\, \lt_{i_1})$. Hence, $\scp(\lt_{i_2\hskip-0.05em}) \prv \scp(\lt_{i_2} |\, \lt_{i_1})$. Therefore, a conditional scope $\scp(\lt_{i_k} |\, \lt_{i_{k-1}})$ can be derived from its scope $\scp(\lt_{i_k\hskip-0.05em})$, viz., $\scp(\lt_{i_k}) \prv \scp(\lt_{i_k} |\, \lt_{i_{k-1}})$. That is, because $\scp(\lt_{i_k\hskip-0.05em})$ is consistent, $\scp(\lt_{i_k} |\, \lt_{i_{k-1}})$ is consistent (cf. \cref{incnstnt}).
\end{proof}

\begin{note}\label{aftrTrmn}
After the scan termination, $\FM$ reduces to  $\FM'\hskip-0.09em$, i.e., $\FM \rdc \FM'\hskip-0.12em$ (\texttt{Scan} L24).  That is,  $\Fm \supseteq \Fm'\hskip-0.09em$, while $\scp \subseteq \scp'\hskip-0.09em$. Let $\FM \gets \FM'\hskip-0.09em$.  Then, $\FM = \scp \wedge \Fm$ such that  $\scp$ and $\Fm$ are \emph{properly} disjoint, in which $\scp$ is consistent.  $\Li$ denotes the literals in $\scp$ and $\Li'\hskip-0.12em$ denotes the literals in  $\Fm$, $\Fm = \hskip-0.09em\bigwedge \hskip-0.09em C_{k\hskip-0.05em}$.
\end{note}

\begin{theorem}[Satisfiability]\label{final}
The following statements are equivalent for any $\{i, j\} \subseteq \Li'\hskip-0.09em$.

$p_1\hskip-0.05em \colon$\emph{Before} the termination, as $\scp(\lt_j)$ was inconsistent,  $\scp  \gets  \scp \wedge \nlt_{j\hskip-0.12em}$ and $\Li \gets \Li \cup \{j\}$, that is, $\lt_{j\hskip-0.15em}$ was  removed from $\Fm$  and $j$ from $\Li'\hskip-0.09em$. Otherwise,  $\lt_{i\hskip-0.07em} \wedge \Fm \rdc \scp(\lt_i) \wedge \Fm'\hskip-0.05em(\lt_i)$, that is, $\scp(\lt_i)$ was consistent such that $\scp(\lt_i)$ and $\Fm'\hskip-0.05em(\lt_i)$ were properly disjoint. Then, it was redundant to check whether  $\unsat \Fm'\hskip-0.05em(\lt_i)$ in order to decide if $\unsat \Fm(\lt_i)$.  Thus, $\unsat \Fm(\lt_j)\hskip-0.11em$ iff $\scp(\lt_j)\hskip-0.09em$ was inconsistent.

$p_2 \colon$\emph{After} the termination, $\Fm \rdc \scp(\lt_i) \wedge \Fm'\hskip-0.05em(\lt_i)$ for each $\lt_{i\hskip-0.11em} \in X_{i\hskip-0.09em}$, where $X_{i\hskip-0.11em} = \{x_i, \nx{i}\}$.

$p_3 \colon \Fm \rdc  \scp(\lt_{i_0}) \wedge  \scp(\lt_{i_1} |\, \lt_{i_0}) \wedge \cdots \wedge \scp(\lt_{i_n} |\, \lt_{i_{n-1}})$, that is, $\Fm$ the formula  is reducible to a  minterm \emph{consistent}, thus $\Fm$ is \emph{satisfiable}. Then, $\alpha$ denotes a satisfying assignment such that $\alpha = \scp(\lt_{i_0}) \cup \scp(\lt_{i_1} |\, \lt_{i_0}) \cup \cdots \cup \scp(\lt_{i_n} |\, \lt_{i_{n-1}})$, in which $\Li(\lt_{i_0}) \cap \Li(\lt_{i_1} |\, \lt_{i_0}) \cap \cdots \cap \Li(\lt_{i_n} |\, \lt_{i_{n-1}}) = \emptyset$.
\end{theorem}
\begin{proof}

The proof is to show that  $p_{1\hskip-0.15em} \imp p_{2\hskip-0.05em}$, $p_{2\hskip-0.09em} \imp p_{3\hskip-0.03em}$, and $p_{3\hskip-0.09em} \imp p_{1\hskip-0.12em}$ (see pg. 88 in \cite{Ros}).

It is obvious that $p_{2\hskip-0.09em} \hskip-0.19em\iff\hskip-0.17em p_{1\hskip-0.15em}$ holds, which  denotes a duality theorem (see pg. 34-36 in \cite{Wig}). That is, $\scp(\lt_i)$ is consistent for any $i \in \Li'\hskip-0.09em$ and \emph{each} $\lt_{i\hskip-0.09em} \in X_{i\hskip-0.05em}$ iff $\scp(\lt_j)$ is inconsistent for any $j \in \ell$ and \emph{some} $\lt_{j\hskip-0.09em} \in X_{j\hskip-0.05em}$, $\ell \subseteq \Li$, after  \texttt{Scan}~$\!(\FM)$ terminates.   In other words,  \emph{each} $\lt_{i\hskip-0.09em} \in X_{i\hskip-0.09em}$ is compatible for any $i \in \Li'\hskip-0.09em$ iff \emph{some} $\lt_{j\hskip-0.09em} \in X_{j\hskip-0.12em}$  is incompatible, thus $\nlt_{j\hskip-0.09em}$ is necessary, for any $j \in \Li$, after  \texttt{Scan}~$\!(\FM)$ terminates. See also \crefrange{NonTrvNss}{CmpL}, as well as  \cref{NonTrv,TRmnt}.

For $p_{2\hskip-0.09em} \imp p_{3\hskip-0.03em}$,  the proof is to show that the construction process of $\hat{\scp}$ the minterm, i.e., $\hat{\scp} = \scp(\lt_{i_0}) \wedge  \scp(\lt_{i_1} |\, \lt_{i_0}) \wedge \cdots \wedge \scp(\lt_{i_n} |\, \lt_{i_{n-1}})$, exhibits the Markov   property, which \emph{preserves} consistency of $\hat{\scp}$. Then, production of the next minterm $\scp(\lt_{i_k} |\, \lt_{i_{k-1}})$  depends \emph{only} upon the current  formula $\Fm'\hskip-0.05em(\lt_{i_{k-1}} |\, \lt_{i_{k-2}})$, that is, it does not depend on the past $\Fm'\hskip-0.05em(\lt_{i_0}),  \,\Fm'\hskip-0.05em(\lt_{i_1} |\, \lt_{i_0}), \ldots, \linebreak \Fm'\hskip-0.05em(\lt_{i_{k-2}} |\, \lt_{i_{k-3}})$. Hence, $ \Fm \rdc \scp(\lt_{i_0}) \wedge \Fm'\hskip-0.05em(\lt_{i_0}), \hskip0.07em \Fm'\hskip-0.05em(\lt_{i_0}) \rdc \scp(\lt_{i_1} |\, \lt_{i_0}) \wedge \Fm'\hskip-0.05em(\lt_{i_1} |\, \lt_{i_0}),  \hskip0.07em \Fm'\hskip-0.05em(\lt_{i_1} |\, \lt_{i_0}) \rdc \scp(\lt_{i_2} |\, \lt_{i_1}) \wedge \Fm'\hskip-0.05em(\lt_{i_2} |\, \lt_{i_1}), \ldots,  \Fm'\hskip-0.05em(\lt_{i_{n-1}} |\, \lt_{i_{n-2}}) \rdc  \scp(\lt_{i_n} |\, \lt_{i_{n-1}})$. Then, appending  $\scp(\lt_{i_n} |\, \lt_{i_{n-1}})$ to $\big(\scp(\lt_{i_0}) \wedge  \scp(\lt_{i_1} |\, \lt_{i_0}) \wedge \cdots \wedge \scp(\lt_{i_{n-1}} |\, \lt_{i_{n-2}})\big)$  \emph{preserves}  consistency of $\hat{\scp}$. Thus, $\Fm \rdc \hat{\scp}$, and $\Fm$ is satisfiable. The  following steps  specify the construction  of  $\hat{\scp}$ (see also  \cref{prdctns,prv}).

Step 0. Follows from the statement $p_{2\hskip-0.11em}$ (see \cref{Scope,Indp,subset}). Pick  $i_0 \hskip-0.09em\in \Li'\hskip-0.09em$ and  $\lt_{i_0\hskip-0.19em} \in X_{i_0\hskip-0.09em}$, thus $\lt_{i_0\hskip-0.05em} \wedge \Fm \rdc \scp(\lt_{i_0}) \wedge \Fm'\hskip-0.05em(\lt_{i_0})$, or $\Fm \rdc \scp(\lt_{i_0}) \wedge \Fm'\hskip-0.05em(\lt_{i_0})$. Then, $\Fm \supseteq \Fm'\hskip-0.05em(\lt_{i_0})$, $\Li'\hskip-0.15em =  \Li(\lt_{i_0}) \cup \Li'\hskip-0.05em(\lt_{i_0})$, and $\Li(\lt_{i_0}) \cap \Li'\hskip-0.05em(\lt_{i_0}) = \emptyset$, i.e., $\scp(\lt_{i_0})$ and $\Fm'\hskip-0.05em(\lt_{i_0})$ are properly disjoint.

Step 1. Pick any $i_1 \hskip-0.15em\in \Li'\hskip-0.05em(\lt_{i_0})$ and  $\lt_{i_1\hskip-0.23em} \in X_{i_1\hskip-0.11em}$, thus $\lt_{i_1\hskip-0.17em} \wedge \Fm'\hskip-0.05em(\lt_{i_0}) \rdc \scp(\lt_{i_1} |\, \lt_{i_0}) \wedge \Fm'\hskip-0.05em(\lt_{i_1} |\, \lt_{i_0})$, or $\Fm'\hskip-0.05em(\lt_{i_0}) \rdc \scp(\lt_{i_1} |\, \lt_{i_0}) \wedge \Fm'\hskip-0.05em(\lt_{i_1} |\, \lt_{i_0})$. That is, $\Fm'\hskip-0.05em(\lt_{i_0}) \supseteq \Fm'\hskip-0.05em(\lt_{i_1} |\, \lt_{i_0})$, and $\scp(\lt_{i_1} |\, \lt_{i_0})$ and $\Fm'\hskip-0.05em(\lt_{i_1} |\, \lt_{i_0})$ are properly disjoint. Then,  $\Li'\hskip-0.05em(\lt_{i_0}) = \Li(\lt_{i_1} |\, \lt_{i_0}) \cup \Li'\hskip-0.05em(\lt_{i_1} |\, \lt_{i_0})$ and $\Li(\lt_{i_1} |\, \lt_{i_0}) \cap \Li'\hskip-0.05em(\lt_{i_1} |\, \lt_{i_0}) = \emptyset$. Also, from step 0, $\Li'\hskip-0.15em =  \Li(\lt_{i_0}) \cup  \Li'\hskip-0.05em(\lt_{i_0})$ and $\Li(\lt_{i_0}) \cap \Li'\hskip-0.05em(\lt_{i_0}) = \emptyset$. Because $\Li'\hskip-0.05em(\lt_{i_0}) \supseteq \Li(\lt_{i_1} |\, \lt_{i_0})$, $\Li(\lt_{i_0}) \cap \Li(\lt_{i_1} |\, \lt_{i_0}) = \emptyset$.  Because $\Li'\hskip-0.05em(\lt_{i_0}) \supseteq \Li'\hskip-0.05em(\lt_{i_1} |\, \lt_{i_0})$, $\Li(\lt_{i_0}) \cap \Li'\hskip-0.05em(\lt_{i_1} |\, \lt_{i_0}) = \emptyset$.  Consequently, $\Li'\hskip-0.15em = \Li(\lt_{i_0}) \cup \Li(\lt_{i_1} |\, \lt_{i_0}) \cup \Li'\hskip-0.05em(\lt_{i_1} |\, \lt_{i_0})$ and $\Li(\lt_{i_0}) \cap \Li(\lt_{i_1} |\, \lt_{i_0}) \cap \Li'\hskip-0.05em(\lt_{i_1} |\, \lt_{i_0}) = \emptyset$. That is, $\Li'\hskip-0.12em$ is \emph{partitioned} into $\Li(\lt_{i_0})$, $\Li(\lt_{i_1} |\, \lt_{i_0})$, and $\Li'\hskip-0.05em(\lt_{i_1} |\, \lt_{i_0})$. Recall that $\Fm \supseteq \Fm'\hskip-0.05em(\lt_{i_0})$ from step 0 and  $\Fm'\hskip-0.05em(\lt_{i_0}) \rdc \scp(\lt_{i_1} |\, \lt_{i_0})$ from step 1. Also, $\Fm \rdc \scp(\lt_{i_1})$ from step 0. Then, $\scp(\lt_{i_1\hskip-0.05em}) \supseteq \scp(\lt_{i_1} |\, \lt_{i_0})$. Hence, $\scp(\lt_{i_1\hskip-0.05em}) \prv \scp(\lt_{i_1} |\, \lt_{i_0})$. Recall that $\scp(\lt_{i_0})$ is consistent and  $\Li(\lt_{i_0}) \cap \Li(\lt_{i_1} |\, \lt_{i_0}) = \emptyset$. As a result, $\big(\scp(\lt_{i_0}) \wedge \scp(\lt_{i_1} |\, \lt_{i_0})\big)$ is consistent. Therefore, $\Fm  \rdc \scp(\lt_{i_0}) \wedge \scp(\lt_{i_1} |\, \lt_{i_0}) \wedge \Fm'\hskip-0.05em(\lt_{i_1} |\, \lt_{i_0})$.

Step 2. Pick any $i_{2\hskip-0.09em} \in \Li'\hskip-0.05em(\lt_{i_1} |\, \lt_{i_0})$ and  $\lt_{i_2\hskip-0.17em} \in X_{i_2\hskip-0.05em}$, thus $ \Fm'\hskip-0.05em(\lt_{i_1} |\, \lt_{i_0}) \hskip-0.05em \rdc \scp(\lt_{i_2} |\, \lt_{i_1\hskip-0.05em}) \wedge \Fm'\hskip-0.05em(\lt_{i_2} |\, \lt_{i_1\hskip-0.05em})$. Then, $\Li'\hskip-0.05em(\lt_{i_1} |\, \lt_{i_0}) = \Li(\lt_{i_2} |\, \lt_{i_1\hskip-0.05em}) \cup \Li'\hskip-0.05em(\lt_{i_2} |\, \lt_{i_1\hskip-0.05em})$ such that $\Li(\lt_{i_2} |\, \lt_{i_1\hskip-0.05em}) \cap \Li'\hskip-0.05em(\lt_{i_2} |\, \lt_{i_1\hskip-0.05em}) = \emptyset$. Also, from step 1, $\Li'\hskip-0.15em = \Li(\lt_{i_0}) \cup \Li(\lt_{i_1} |\, \lt_{i_0}) \cup \Li'\hskip-0.05em(\lt_{i_1} |\, \lt_{i_0})$ such that $\Li(\lt_{i_0}) \cap \Li(\lt_{i_1} |\, \lt_{i_0}) \cap \Li'\hskip-0.05em(\lt_{i_1} |\, \lt_{i_0}) = \emptyset$. As a result, $\Li(\lt_{i_0}) \cap \Li(\lt_{i_1} |\, \lt_{i_0}) \cap \Li(\lt_{i_2} |\, \lt_{i_1\hskip-0.05em}) = \emptyset$. Hence, $\big(\Li(\lt_{i_0}) \cup \Li(\lt_{i_1} |\, \lt_{i_0})\big) \cap \Li(\lt_{i_2} |\, \lt_{i_1\hskip-0.05em}) = \emptyset$.  Note that $\scp(\lt_{i_2\hskip-0.05em}) \prv   \scp(\lt_{i_2} |\, \lt_{i_1\hskip-0.05em})$. Consequently,  appending $\scp(\lt_{i_2} |\, \lt_{i_1\hskip-0.05em})$ to $\big(\scp(\lt_{i_0}) \wedge \scp(\lt_{i_1} |\, \lt_{i_0})\big)$ preserves the consistency. That is, $\big(\scp(\lt_{i_0}) \wedge \scp(\lt_{i_1} |\, \lt_{i_0}) \wedge  \scp(\lt_{i_2} |\, \lt_{i_1\hskip-0.05em})\big)$  is consistent. Therefore, $\Fm  \rdc \scp(\lt_{i_0}) \wedge \scp(\lt_{i_1} |\, \lt_{i_0}) \wedge  \scp(\lt_{i_2} |\, \lt_{i_1\hskip-0.05em}) \wedge \Fm'\hskip-0.05em(\lt_{i_2} |\, \lt_{i_1\hskip-0.05em})$.

Step 3.   $\Li'\hskip-0.12em$ is partitioned into $\big(\Li(\lt_{i_0}) \cup \Li(\lt_{i_1} |\, \lt_{i_0}) \cup \Li(\lt_{i_2} |\, \lt_{i_1\hskip-0.05em})\big)$,  $\Li(\lt_{i_3} |\, \lt_{i_2\hskip-0.05em})$, and $\Li'\hskip-0.05em(\lt_{i_3} |\, \lt_{i_2\hskip-0.05em})$.

Step $n$. $\Li'\hskip-0.05em(\lt_{i_n} |\, \lt_{i_{n-1}}) = \emptyset$. $\Li'\hskip-0.12em$ is partitioned into $\big(\Li(\lt_{i_0}) \cup \Li(\lt_{i_1} |\, \lt_{i_0}) \cup  \cdots \cup \Li(\lt_{i_{n-1}} |\, \lt_{i_{n-2}})\big)$ and $\Li(\lt_{i_n} |\, \lt_{i_{n-1}})$. That is, $\big(\scp(\lt_{i_0}) \wedge \scp(\lt_{i_1} |\, \lt_{i_0}) \wedge  \cdots \wedge \scp(\lt_{i_{n-1}} |\, \lt_{i_{n-2}})\big)$ and $\scp(\lt_{i_n} |\, \lt_{i_{n-1}})$ are properly disjoint, which are consistent as well. Thus, $\hat{\scp} =  \scp(\lt_{i_0}) \wedge  \bigwedge_{k = 1}^n \scp(\lt_{i_k} |\, \lt_{i_{k-1}})$, which is \emph{consistent}. Since $\Fm \rdc \hat{\scp}$,  $\Fm$ is \emph{satisfiable} and $\sat \Fm$ by $\alpha = \scp(\lt_{i_0}) \cup  \bigcup_{k = 1}^n \scp(\lt_{i_k} |\, \lt_{i_{k-1}})$, which denotes a satisfying assignment. Recall that $\Fm \gets \Fm'\hskip-0.15em$ (see \cref{aftrTrmn}). Therefore, $p_{2\hskip-0.05em} \imp p_{3\hskip-0.05em}$ holds.

Finally, we show $p_{3\hskip-0.09em} \imp p_{1\hskip-0.09em}$.  $ \Fm  \supseteq \Fm'\hskip-0.05em(\lt_i)$  (from the statement $p_2$).  $\Fm$ is satisfiable (from step $n$). Thus,  $\Fm'\hskip-0.05em(\lt_i)$ becomes satisfiable, \emph{after}   \texttt{Scan} terminates. As a result, it is \emph{redundant} to check unsatisfiability of $\Fm'\hskip-0.05em(\lt_i)$  in order to decide unsatisfiability of $\Fm(\lt_i)$, where $\Fm(\lt_i) = \scp(\lt_i) \wedge \Fm'\hskip-0.05em(\lt_i)$ such that $\scp(\lt_i)$ and $\Fm'\hskip-0.05em(\lt_i)$ are properly disjoint.  Therefore, inconsistency of  $\scp(\lt_j)$ the minterm, which is sufficient, becomes  necessary also for the unsatisfiability of $\Fm(\lt_j)$, thus $\unsat \Fm(\lt_j)$ iff $\scp(\lt_j)$ is inconsistent, \emph{before}  \texttt{Scan} terminates. See also  \cref{incnstnt,assmp}.
\end{proof}

\begin{corollary}[Prime Normal Form]\label{PnF}
$\Psi \hskip-0.07em=\hskip-0.07em  \bigwedge_{i \in \Li'} \hskip-0.19em \big(\scp(x_i) \vee \scp(\nx i)\big)$, as $\lt_{i\hskip-0.09em} \rdc \scp(\lt_i)$ for all $\lt_{i\hskip-0.11em} \in X_{i\hskip-0.09em}$.
\end{corollary}
\begin{proof}
Follows  from \cref{final}. Note that $\Psi$ denotes the semantics and $\Fm$ the syntax.
\end{proof}

\begin{proposition}
The complexity of X3SAT is $O(mn^3)$, and of XSAT is $O(mn^4)$.
\end{proposition}
\begin{proof}
The proof is obvious (see \hyperref[alg:Scan]{\texttt{Scan}}). Note that $|C_k| \leqslant 3$, or $|C_k| \leqslant n$  by \cref{clause}.
\end{proof}

\subsection[Assignment]{Construction of a Satisfying Assignment}\label{s:Assg}

Let $(i_0, i_{1\hskip-0.05em}, \ldots, i_n)$ be a literal ordering such that $i_0\hskip-0.07em \in \Li'\hskip-0.09em, i_{1\hskip-0.15em} \in \Li'\hskip-0.05em(\lt_{i_0}),  \ldots, i_n\hskip-0.12em \in \Li'\hskip-0.05em(\lt_{i_{n-1}} |\, \lt_{i_{n-2}})$. Then, $\alpha = \scp(\lt_{i_0}) \cup \scp(\lt_{i_1\hskip-0.05em}) \cup \cdots \cup \scp(\lt_{i_n})$. Also, $\alpha = \scp(\lt_{i_0}) \cup \scp(\lt_{i_1} |\, \lt_{i_0}) \cup \cdots \cup \scp(\lt_{i_n} |\, \lt_{i_{n-1}})$, in which $\scp(\lt_{i_0}), \scp(\lt_{i_1} |\, \lt_{i_0}),  \ldots, \scp(\lt_{i_n} |\, \lt_{i_{n-1}})$ are properly disjoint (see \cref{PrpDsj}), where $\scp(\lt_{i_1} |\, \lt_{i_0}) \hskip-0.05em= \scp(\lt_{i_1\hskip-0.05em}) - \scp(\lt_{i_0})$ and $\scp(\lt_{i_k} |\, \lt_{i_{k-1}}) = \scp(\lt_{i_k}) - \big(\scp(\lt_{i_0}) \cup \bigcup_{l = 2}^k \scp(\lt_{i_{l-1}} |\, \lt_{i_{l-2}})\big)$.

\begin{example}
Let $\Fm = (x_{1\hskip-0.12em} \X \nx{2\hskip-0.05em} \X x_6) \wedge (x_{3\hskip-0.05em} \X x_{4\hskip-0.05em} \X \nx5) \wedge (x_{3\hskip-0.05em} \X x_{6\hskip-0.05em} \X \nx7) \wedge (x_{4\hskip-0.05em} \X x_{6\hskip-0.05em} \X \nx7)$. Then,   $\scp = \nx{3\hskip-0.05em} \wedge \nx{4\hskip-0.05em} \wedge \nx{5\hskip-0.05em}$ and $\Fm'\hskip-0.15em = (x_{1\hskip-0.12em} \X \nx{2\hskip-0.05em} \X x_6) \wedge (x_{6\hskip-0.05em} \X \nx7)$. Thus,  $\Li'\hskip-0.15em = \{1, 2, 6, 7\}$. Consider the ordering $(\nx{7\hskip-0.03em}, x_2, x_1)$. Note that $7 \in \Li'\hskip-0.09em$, $2 \in \Li'\hskip-0.05em(\nx7)$, and $1 \in \Li'\hskip-0.05em(x_2 |\, \nx7)$. Then, $\sat \Fm'\hskip-0.09em$ by $\alpha = \scp(\nx7) \cup  \scp(x_2) \cup  \scp(x_1)$, where $\scp(\nx7) = \{\nx{7\hskip-0.03em}, \nx6\}$, $\scp(x_2) = \{x_2\}$, and $\scp(x_1) = \{x_1, x_2, \nx{7\hskip-0.03em}, \nx6\}$. Also, $(\nx{7\hskip-0.03em}, x_2, x_1) \rdc  \scp(\nx7) \wedge \scp(x_2 |\, \nx7) \wedge \scp(x_1 |\, x_2)$, in which  $\scp(x_2 |\, \nx7) =  \scp(x_2) - \scp(\nx7)$, and  $\scp(x_1 |\, x_2) = \scp(x_1) -  \big(\scp(x_2 |\, \nx7) \cup \scp(\nx7)  \big)$. Then, $\sat \Fm'\hskip-0.09em$ by $\alpha = \{\nx{7\hskip-0.03em}, \nx6\} \cup \{x_2\} \cup \{x_1\}$.  That is, $x_{7\hskip-0.07em} = 0, x_{6\hskip-0.07em} = 0, x_{2\hskip-0.07em} = 1, x_{1\hskip-0.11em} = 1$ for satisfying $\Fm'\hskip-0.09em$. Also, $x_{3\hskip-0.07em} = 0, x_{4\hskip-0.07em} = 0, x_{5\hskip-0.07em} = 0$ by $\scp$,  fixed for  $\Fm$.
\end{example}

\subsection[QBF]{Quantified Propositional Logic: \textit{TQBF}}\label{s:QxSAT}
\textit{TQBF} is \textsf{\bf PSPACE}-complete (see pg. 339 in \cite{Sip}). Let  $\fm =  \hskip-0.09em\bigwedge\hskip-0.09em c_{k\hskip-0.09em}$ be a 3SAT formula, where $c_{k\hskip-0.11em} = \linebreak (\lt_{i\hskip-0.07em} \vee \lt_{j\hskip-0.09em} \vee \lt_u)$. Let $\Fm = \hskip-0.09em\bigwedge\hskip-0.09em C_{k\hskip-0.09em}$ be an X3SAT formula transformed from $\fm$. Let $\Fm$ be satisfiable.

\begin{definition}[Quantified Boolean Formula]\label{QBF}
$Q_1 \lt_{1\hskip-0.03em} \, Q_2 \lt_2 \cdots Q_n \lt_n \,\fm$,  where $Q_{i\hskip-0.09em} \in \{\exists, \forall\}$.
\end{definition}

\begin{note*}
The QBF in  \cref{QBF} is conventionally expressed by $Q_1 x_{1\hskip-0.03em} \, Q_2 x_2  \cdots Q_n x_n \,\fm$, in which $x_{i\hskip-0.09em} \in \{0 ,1\}$. Thus, $\lt_{i\hskip-0.09em} = x_{i\hskip-0.09em}$ iff $x_{i\hskip-0.09em} = 1$, and $\lt_{i\hskip-0.09em} = \nx{i\hskip-0.09em}$ iff $x_{i\hskip-0.09em} = 0$ iff $\nx{i\hskip-0.09em} = 1$, since $\lt_{i\hskip-0.09em} \in \{x_i, \nx{i}\}$.
\end{note*}

Firstly, prime assignments of each clause $c_{k\hskip-0.09em}$ are determined, which are denoted by $\scp_k^i \hskip-0.03em$. Next, the Prime Normal Form (PNF) is constructed  based on the prime assignments.

\begin{definition}[Prime assignments]\label{PA}
$\lt_{i\hskip-0.07em} \wedge \nlt_{j\hskip-0.05em}$,  $\nlt_{i\hskip-0.07em} \wedge \lt_{j\hskip-0.15em}$ and $\lt_{i\hskip-0.07em} \wedge \lt_{j\hskip-0.15em}$ are the prime assignments for  $c_{k\hskip-0.12em} = (\lt_{i\hskip-0.13em} \vee \lt_j)$, in which $\scp_k^{_1} \hskip-0.09em = \lt_{i\hskip-0.07em} \wedge \nlt_{j\hskip-0.05em}$. Likewise, $\lt_{i\hskip-0.05em} \wedge \nlt_{j\hskip-0.07em} \wedge \nlt_{u\hskip-0.05em}$, $\nlt_{i\hskip-0.05em} \wedge \lt_{j\hskip-0.07em} \wedge \nlt_{u\hskip-0.05em}, \ldots,  \lt_{i\hskip-0.05em} \wedge \lt_{j\hskip-0.07em} \wedge \lt_{u\hskip-0.15em}$ are the prime assignments for  $c_{k'\hskip-0.12em} = (\lt_{i\hskip-0.15em} \vee \lt_{j\hskip-0.15em} \vee \lt_u)$,  in which $\scp_{k'}^{_7} \hskip-0.09em = \lt_{i\hskip-0.07em} \wedge \lt_{j\hskip-0.09em} \wedge \lt_{u\hskip-0.05em}$.
\end{definition}

\begin{definition}[Prime clause]\label{PC}
$\delta_{\hskip-0.05em k\hskip-0.12em}$ is a disjunction of prime assignments such that either $\delta_{\hskip-0.05em k\hskip-0.12em} = \linebreak  (\scp_k^{_1\hskip-0.19em} \vee \scp_k^{_2\hskip-0.09em} \vee \scp_k^{_3})$, or $\delta_{\hskip-0.05em k\hskip-0.12em} = (\scp_k^{_1\hskip-0.19em} \vee \scp_k^{_2\hskip-0.09em} \vee \cdots \vee \scp_k^{_7})$. Thus, $\delta_{\hskip-0.05em k\hskip-0.12em}$ denotes a Disjunction Normal Form.
\end{definition}

\begin{definition}[PNF]\label{PNF}
$\Psi =  \bigwedge_{k=1}^m \hskip-0.12em \bigvee_{i=1}^n \hskip-0.05em \scp_k^i\hskip-0.12em$ such that $\scp_k^i \hskip-0.09em\wedge \fm$ is satisfiable,  $n \in \{1, 2, \ldots, 7\}$.
\end{definition}

Note that $\scp_k^i \hskip-0.09em\wedge \fm$ is satisfiable iff $\scp_k^i \hskip-0.09em\wedge \Fm$ is satisfiable, because  $\fm$ and $\Fm$  are equisatisfiable. Note also that $\fm$ denotes the syntax and $\Psi$ denotes the semantics (cf. \cref{PnF}).

\begin{lemma}[Collapse of a prime clause to a prime satisfying assignment]\label{CkCllps}
$\scp_k^i\hskip-0.07em \wedge \delta_{\hskip-0.05em k\hskip-0.09em} \prv \scp_k^i\hskip-0.03em$.
\end{lemma}
\begin{proof}
Follows directly from \crefrange{PA}{PNF}.
\end{proof}

\begin{note}\label{SatAssg}
Because $\fm$ is satisfiable, there exists  $(\scp_1^i\hskip-0.11em \wedge \scp_2^j\hskip-0.07em \wedge \cdots \wedge \scp_m^u)$ consistent with $\Psi$.
\end{note}

\begin{definition}[Legal moves]\label{LegalMoves}
$\tilde{\fm} = \hskip-0.09em\bigwedge \hskip-0.09em \tilde{c}_{k\hskip-0.05em}$, a 2SAT formula, in which $\tilde{c}_{k\hskip-0.11em} = \nlt_{i\hskip-0.11em} \vee \nlt_{u\hskip-0.17em}$ such that $\lt_{i\hskip-0.05em} \wedge \lt_{u\hskip-0.07em} \wedge \Psi$ is inconsistent or  $\lt_{i\hskip-0.05em} \wedge \lt_{u\hskip-0.07em} \wedge \fm$ leads to a conjunct $\lt_{j\hskip-0.12em}$ for some $j \in A$, $A = \{2, 4, \ldots, n\}$, for any $(i, u)$ in $\{(1, 3), (1, 5), \ldots, (1, n-1), (3, 5),  \ldots, (3, n-1), (5, 7), \ldots, (n-3, n-1) \}$.
\end{definition}

\begin{theorem}[True QBF]\label{TQBF}
$\exists \lt_1 \forall \lt_2 \, \exists \lt_{3\hskip-0.03em} \, \forall \lt_4 \cdots \exists \lt_{n-1\hskip-0.03em} \forall \lt_n \, \fm$ is true iff the following statement is true, in which  $c.\scp(\lt_i, \lt_j)$ denotes that $\scp(\lt_i, \lt_j) \wedge \Psi \wedge \tilde{\fm}$ is consistent.

$c.\scp(x_{1\hskip-0.05em}, x_2)$ \tabto{7.8em} and \tabto{10em} $c.\scp(x_{1\hskip-0.05em}, \nx2)$  \tabto{16.5em} OR \tabto{19.5em} $c.\scp(\nx{1\hskip-0.05em}, x_2)$ \tabto{26em} and \tabto{28.2em} $c.\scp(\nx{1\hskip-0.05em}, \nx2)$ \tabto{35em} AND $\quad 1$

$c.\scp(x_{1\hskip-0.05em}, x_4)$ \tabto{7.8em} and \tabto{10em} $c.\scp(x_{1\hskip-0.05em}, \nx4)$  \tabto{16.5em} OR \tabto{19.5em} $c.\scp(\nx{1\hskip-0.05em}, x_4)$ \tabto{26em} and \tabto{28.2em} $c.\scp(\nx{1\hskip-0.05em}, \nx4)$ \tabto{35em} AND

 \vskip-0.1em
 \tabto{17.1em} $\vdots$
 \vskip-0.3em

$c.\scp(x_{1\hskip-0.05em}, x_n)$ \tabto{7.8em} and \tabto{10em} $c.\scp(x_{1\hskip-0.05em}, \nx n)$  \tabto{16.5em} OR \tabto{19.5em} $c.\scp(\nx{1\hskip-0.05em}, x_n)$ \tabto{26em} and \tabto{28.2em} $c.\scp(\nx{1\hskip-0.05em}, \nx n)$ \tabto{35em} AND

$c.\scp(x_2, x_3)$ \tabto{7.8em} or \tabto{10em} $c.\scp(x_2, \nx3)$   \tabto{16.5em} AND \tabto{19.5em} $c.\scp(\nx2, x_3)$ \tabto{26em} or \tabto{28.2em} $c.\scp(\nx2, \nx3)$ \tabto{35em} AND $\quad 4$

$c.\scp(x_2, x_5)$ \tabto{7.8em} or \tabto{10em} $c.\scp(x_2, \nx5)$  \tabto{16.5em} AND \tabto{19.5em} $c.\scp(\nx2, x_5)$ \tabto{26em} or \tabto{28.2em} $c.\scp(\nx2, \nx5)$ \tabto{35em} AND

 \vskip-0.1em
 \tabto{17.1em} $\vdots$
 \vskip-0.3em

$c.\scp(x_2, x_{n-1})$ \tabto{7.8em} or \tabto{10em} $c.\scp(\nx2, x_{n-1})$ \tabto{16.5em} AND \tabto{19.5em} $c.\scp(x_2, \nx{n-1})$ \tabto{26em} or \tabto{28.2em} $c.\scp(\nx2, \nx{n-1})$ \tabto{35em} AND

$c.\scp(x_3, x_4)$ \tabto{7.8em} and \tabto{10em} $c.\scp(x_3, \nx4)$  \tabto{16.5em} OR \tabto{19.5em} $c.\scp(\nx3, x_4)$ \tabto{26em} and \tabto{28.2em} $c.\scp(\nx3, \nx4)$ \tabto{35em} AND

$c.\scp(x_3, x_6)$ \tabto{7.8em} and \tabto{10em} $c.\scp(x_3, \nx6)$  \tabto{16.5em} OR \tabto{19.5em} $c.\scp(\nx3, x_6)$ \tabto{26em} and \tabto{28.2em} $c.\scp(\nx3, \nx6)$ \tabto{35em} AND

 \vskip-0.1em
 \tabto{17.1em} $\vdots$
 \vskip-0.3em

$c.\scp(x_{n-1}, x_n)$ \tabto{7.8em} and \tabto{10em} $c.\scp(x_{n-1}, \nx n)$ \tabto{16.5em} OR \tabto{19.5em} $c.\scp(\nx{n-1}, x_n)$ \tabto{26em} and \tabto{28.2em} $c.\scp(\nx{n-1}, \nx n)$.
\end{theorem}
\begin{proof}[Proof sketch]
Let $\Phi \hskip-0.11em\coloneqq \Psi \wedge \tilde{\fm}$, $\Phi(\lt_i, \lt_j) \hskip-0.11em\coloneqq \scp(\lt_i, \lt_j) \wedge \Phi$, and $\scp(\lt_i, \lt_j) \hskip-0.11em\coloneqq \lt_{i\hskip-0.05em} \wedge \lt_{j\hskip-0.05em}$. Recall that $A = \{2, 4, \ldots, n\}$, which denotes the universally quantified literals. Consider the evaluation of $\Phi(x_{1\hskip-0.05em}, x_2)$. $\nlt_{i\hskip-0.09em}$  is removed from any $\tilde{c}_{k\hskip-0.09em}$ such that $  \lt_{i\hskip-0.09em} \in \scp(x_{1\hskip-0.05em}, x_2)$, thus $\scp(x_{1\hskip-0.05em}, x_2) \gets \scp(x_{1\hskip-0.05em}, x_2) \wedge \tilde{c}_{k\hskip-0.09em}$ and any $\tilde{c}_{k\hskip-0.09em}$ is removed from $\tilde{\fm}$. Every $\delta_{\hskip-0.05em k\hskip-0.09em}$ is removed from $\Psi$ such that $\scp_k^i\hskip-0.11em \subseteq \scp(x_{1\hskip-0.05em}, x_2)$ (see also  \cref{CkCllps}). Every $\scp_k^i\hskip-0.13em$ containing $\scp_k^j\hskip-0.13em$ is removed from any $\delta_{\hskip-0.05em k\hskip-0.09em}$ such that $\scp(x_{1\hskip-0.05em}, x_2) \wedge \scp_k^j\hskip-0.13em$ is inconsistent. If $\delta_{\hskip-0.05em k\hskip-0.09em} = \scp_k^{u\hskip-0.11em}$ also, then $\scp(x_{1\hskip-0.05em}, x_2) \gets \scp(x_{1\hskip-0.05em}, x_2) \wedge \scp_k^{u\hskip-0.05em}$, and $\delta_{\hskip-0.05em k\hskip-0.09em}$ is removed from $\Psi$. If $\scp(x_{1\hskip-0.05em}, x_2)$ becomes inconsistent, or $\lt_{j\hskip-0.12em} \in \scp(x_{1\hskip-0.05em}, x_2)$ for some $j \in (A - \{2\})$, then ``$c.\scp(\nx1, x_2)$ and $c.\scp(\nx1, \nx2)$'' holds (cf. L1). As a result, $\scp \wedge \forall \lt_2 \, \exists \lt_3 \cdots \exists \lt_{n-1\hskip-0.03em} \forall \lt_n \Phi$ holds, where $\scp \gets \scp \wedge \nx{1\hskip-0.05em}$. Therefore,  $x_{1\hskip-0.15em}$ is removed from each $\tilde{c}_{k\hskip-0.05em}$, $\scp \gets \scp \wedge \tilde{c}_{k\hskip-0.05em}$, and $\tilde{c}_{k\hskip-0.09em}$ is removed from $\tilde{\fm}$. Also, any  $\scp_k^i\hskip-0.13em$ containing $x_{1\hskip-0.15em}$ is removed from each $\delta_{\hskip-0.05em k\hskip-0.05em}$. If $\delta_{\hskip-0.05em k\hskip-0.09em} = \scp_k^{u\hskip-0.11em}$ also, then $\scp \gets \scp \wedge \scp_k^{u\hskip-0.05em}$, and $\delta_{\hskip-0.05em k\hskip-0.09em}$ is removed from $\Psi$. Consider the evaluation of $\Phi(x_2, \nx3)$. If $\scp(x_2, \nx3)$ is inconsistent, or $\lt_{j\hskip-0.12em} \in \scp(x_2, \nx3)$ for some $j \in (A - \{2\})$, then ``$c.\scp(x_2, x_3)$ AND $c.\scp(\nx2, x_3)$ or $c.\scp(\nx2, \nx3)$'' holds (cf. L4), i.e., $x_{2\hskip-0.07em} \imp x_{3\hskip-0.03em}$. Thus, $\Phi \gets \Phi \wedge (\nx{2\hskip-0.11em} \vee x_3)$. If some $\scp(\lt_u, \lt_v)$ becomes inconsistent, or $\lt_{j\hskip-0.12em} \in \scp(\lt_u, \lt_v)$ for some $j \in (A - \{v\})$, then each $\Phi(\lt_i, \lt_j)$ is re-evaluated. Consequently, if $\scp$ or $\Phi$ becomes inconsistent, or if $\lt_{j\hskip-0.12em} \in \scp$ for some $j \in A$, then the QBF is false. Otherwise,  the QBF is true. In this case, $\Phi(\lt_i, \lt_j) = \scp(\lt_i, \lt_j) \wedge \Phi'\hskip-0.05em(\lt_i, \lt_j)$, in which  $\Phi'\hskip-0.05em(\lt_i, \lt_j)$ is contained by $\Phi$. Note that there exists $\big(\scp(\lt_i, \lt_j) \wedge \scp(\lt_j, \lt_k) \wedge \cdots \wedge \scp(\lt_u, \lt_v)\big)$ consistent with $\Phi$ (see also \cref{SatAssg}).
\end{proof}

\begin{note}\label{QBFf}
The QBF is false if $\lt_{j\hskip-0.12em} \in \scp'\hskip-0.12em$ for some $j \in A$ when \texttt{Scan}~$\!(\FM)$ terminates due to L24.
\end{note}

\begin{note*}
The QBF is false if $\fm$ contains a clause $(\lt_{i\hskip-0.07em} \vee \lt_{j\hskip-0.09em} \vee \lt_u)$ such that $\{i, j, u\} \subseteq A$. Note that $\nlt_{i\hskip-0.07em} \wedge \nlt_{j\hskip-0.09em} \wedge \nlt_{u\hskip-0.09em} \wedge \fm$ is unsatisfiable. Recall that $A$ denotes the universally quantified literals.
\end{note*}

\begin{remark}\label{tQBF}
Let $\hat{\Phi}$ be constructed by removing $\{\lt_1, \lt_3, \ldots, \lt_{n-1}\}$ from $\Phi$  if the QBF is true. Then, $\hat{\Phi}$ is \emph{valid}. Recall that \emph{any} $\lt_{i\hskip-0.09em} \in \{x_i, \nx{i}\}$ is \emph{compatible} for all $i \in \Li'\hskip-0.09em$ by \cref{final}.
\end{remark}

\begin{example}
Let $\fm = (0 \vee x_{3\hskip-0.11em} \vee \nx2) \wedge (\nx{1\hskip-0.13em} \vee 1 \vee \nx3)$. Then, $\scp^{_1\!} = x_{3\hskip-0.05em} \wedge x_2$, $\scp^{_2\!} = \nx{3\hskip-0.05em} \wedge \nx2$, and $\scp^{_3\!} = x_{3\hskip-0.05em} \wedge \nx2$. Also, $\scp^{_4\!} = \nx{1\hskip-0.11em}  \wedge x_3$, $\scp^{_5\!} = x_{1\hskip-0.11em}  \wedge \nx3$, $\scp^{_6\!} = \nx{1\hskip-0.11em}  \wedge \nx3$, and $\scp^{_7\!} = x_{1\hskip-0.11em}  \wedge x_3$. Hence, $\Psi =  \linebreak (\scp^{_1\hskip-0.15em} \vee \scp^{_2\hskip-0.09em} \vee \scp^{_3}) \wedge (\scp^{_4\hskip-0.09em} \vee \cdots \vee \scp^{_7})$, in which $\scp^{_1\!} \wedge \scp^{_7\!}$ denotes a satisfying assignment (see \cref{SatAssg}).
\end{example}

\begin{example}
Let $\fm = (x_{1\hskip-0.15em} \vee x_2) \wedge (x_{2\hskip-0.11em} \vee x_3) \wedge (x_{2\hskip-0.11em} \vee \nx3)$. Consider $\exists \lt_{1\hskip-0.03em} \forall \lt_2 \, \exists \lt_3 \, \fm\hskip-0.05em$ (cf. Example 8.10 on pg. 342 in \cite{Sip}). Then, $\tilde{\Psi} = [(x_{1\hskip-0.11em} \wedge x_2) \vee (x_{1\hskip-0.11em} \wedge \nx2) \vee (\nx{1\hskip-0.11em} \wedge x_2)] \wedge [(x_{2\hskip-0.05em} \wedge x_3) \vee (x_{2\hskip-0.05em} \wedge \nx3) \vee (\nx{2\hskip-0.05em} \wedge x_3)] \wedge [(x_{2\hskip-0.05em} \wedge \nx3) \vee (x_{2\hskip-0.05em} \wedge x_3) \vee (\nx{2\hskip-0.05em} \wedge \nx3)]$ by  \cref{PA}. Consider $(x_{1\hskip-0.11em} \wedge \nx2) \wedge \tilde{\Psi}$. Since $(\nx{2\hskip-0.05em} \wedge x_3) \wedge  (\nx{2\hskip-0.05em} \wedge \nx3)$ is inconsistent, $(x_{1\hskip-0.11em} \wedge \nx2)$ is removed from $\tilde{\Psi}$. Also, $(\nx{2\hskip-0.05em} \wedge \lt_3) \wedge \tilde{\Psi}$ leads to inconsistency for each $\lt_{3\hskip-0.11em} \in X_{3\hskip-0.03em}$. Hence, $\Psi = [(x_{1\hskip-0.11em} \wedge x_2) \vee (\nx{1\hskip-0.11em} \wedge x_2)] \wedge [(x_{2\hskip-0.05em} \wedge x_3) \vee (x_{2\hskip-0.05em} \wedge \nx3)]$, in which $\delta_{2\hskip-0.09em} =  \delta_{3\hskip-0.03em}$. Because $2 \in A$ and $\nx{2\hskip-0.07em} \notin \scp^{i\hskip-0.11em}$ for any $i$, the QBF is false. Note that $\nx{2\hskip-0.07em}$ is already incompatible for $\Fm$ by \cref{noScope}. Hence, $x_{2\hskip-0.09em} \in \scp'\hskip-0.09em$, thus the QBF is  false (see \cref{QBFf}).
\end{example}

\begin{example}
$\Psi$ is constructed as in \cref{t:QBF} for $c_{1\hskip-0.2em} = (x_{1\hskip-0.15em} \vee x_{3\hskip-0.13em} \vee \nx2)$ and $c_{2\hskip-0.11em} = (\nx{1\hskip-0.15em} \vee x_{2\hskip-0.13em} \vee \nx4)$.

\begin{table} [!h] \centering \vskip-0.1em
\caption{$\fm = c_{1\hskip-0.09em} \wedge c_{2\hskip-0.07em}$ and $\Psi = \delta_{1\hskip-0.09em} \wedge \delta_{2\hskip-0.03em}$, where  $\delta_{1\hskip-0.15em} = (\scp^{_1\hskip-0.15em} \vee \scp^{_2\hskip-0.09em} \vee \cdots \vee \scp^{_7})$ and $\delta_{2\hskip-0.13em} = (\scp^{_8\hskip-0.09em} \vee \scp^{_9\hskip-0.09em} \vee \cdots \vee \scp^{_{14}})$}\label{t:QBF}
    \begin{tabular}{ r r | r r }
      \hline
        $\scp^{_1\!} = (x_{1\hskip-0.11em} \wedge \nx{3\hskip-0.07em} \wedge x_2)$ & $\scp^{_4\!} = (x_{1\hskip-0.11em} \wedge x_{3\hskip-0.07em} \wedge x_2)\quad$ & $\;\,\scp^{_8\!} = (\nx{1\hskip-0.11em} \wedge \nx{2\hskip-0.07em} \wedge x_4)$ & $\scp^{_{11\!}} = (\nx{1\hskip-0.11em} \wedge x_{2\hskip-0.07em} \wedge x_4)$ \\
        $\scp^{_2\!} = (\nx{1\hskip-0.11em} \wedge x_{3\hskip-0.07em} \wedge x_2)$ & $\scp^{_5\!} = (x_{1\hskip-0.11em} \wedge \nx{3\hskip-0.07em} \wedge \nx2)\quad$ & $\;\,\scp^{_9\!} = (x_{1\hskip-0.11em} \wedge x_{2\hskip-0.07em} \wedge x_4)$ & $\scp^{_{12\!}} = (\nx{1\hskip-0.11em} \wedge \nx{2\hskip-0.07em} \wedge \nx4)$ \\
        $\scp^{_3\!} = (\nx{1\hskip-0.11em} \wedge \nx{3\hskip-0.07em} \wedge \nx2)$ & $\scp^{_6\!} = (\nx{1\hskip-0.11em} \wedge x_{3\hskip-0.07em} \wedge \nx2)\quad$ & $\;\,\scp^{_{10\!}} = (x_{1\hskip-0.11em} \wedge \nx{2\hskip-0.07em} \wedge \nx4)$ & $\scp^{_{13\!}} = (x_{1\hskip-0.11em} \wedge x_{2\hskip-0.07em} \wedge \nx4)$ \\
        & $\scp^{_7\!} = (x_{1\hskip-0.09em} \wedge x_{3\hskip-0.05em} \wedge \nx2)\quad$ & & $\;\scp^{_{14\!}} = (\nx{1\hskip-0.09em} \wedge x_{2\hskip-0.05em} \wedge \nx4)$ \\
      \hline
    \end{tabular} \vskip-0.1em
\end{table}

\noindent Consider  $\exists \lt_{1\hskip-0.03em} \forall \lt_2 \hskip0.07em \exists \lt_3 \forall \lt_4 \hskip0.09em \fm$. Firstly, $\tilde{\fm}$ is determined by \cref{LegalMoves}: $\lt_{1\hskip-0.09em} \wedge \lt_{3\hskip-0.05em} \wedge \Psi$ is consistent for any $\lt_{1\hskip-0.11em} \in X_{1\hskip-0.09em}$ and $\lt_{3\hskip-0.09em} \in X_{3\hskip-0.03em}$, while $\color{red}  \nx{1\hskip-0.11em} \wedge \nx{3\hskip-0.07em} \wedge \fm \prv \nx{2\hskip-0.05em}$. Hence, $\tilde{\fm} = \color{red} x_{1\hskip-0.17em} \vee x_{3\hskip-0.03em}$.  Then, $\Phi(x_{1\hskip-0.03em}, \nx2) \hskip-0.05em=  \scp(x_{1\hskip-0.03em}, \nx2) \wedge  \Phi'\hskip-0.05em(x_{1\hskip-0.03em}, \nx2)$, where $\scp(x_{1\hskip-0.03em}, \nx2) \hskip-0.05em= x_{1\hskip-0.11em} \wedge \nx{2\hskip-0.05em}  \wedge \scp^{_{10}\hskip-0.11em}$ and $\Phi'\hskip-0.05em(x_{1\hskip-0.03em}, \nx2) \hskip-0.05em= (\scp^{_5\hskip-0.15em} \vee \scp^{_7}) \wedge \tilde{\fm}$. Since $\nx{4\hskip-0.12em} \in \scp(x_{1\hskip-0.03em}, \nx2)$,  $x_{1\hskip-0.17em}$ is incompatible, i.e.,  $\nx{1\hskip-0.17em}$ is necessary, thus $\Phi \hskip-0.05em\gets \nx{1\hskip-0.09em} \wedge \Phi$, and $\Phi \hskip-0.05em=  \color{blue}\nx{1\hskip-0.09em} \wedge x_{3\hskip-0.05em} \color{black}  \wedge \Psi$. Note that $\nx{1\hskip-0.15em}$ in $c_{2\hskip-0.09em}$ is already decided to be necessary for $\Phi$, since $\Li(c_2) - A = \{1\}$. Thus,  the QBF is true  by \cref{TQBF}, that is, $\forall x_2 \hskip0.07em \forall x_4 \hskip0.07em [(x_{3\hskip-0.13em} \vee \nx2) \wedge (\nx{1\hskip-0.15em} \vee x_{2\hskip-0.13em} \vee \nx4)]$ is true for $\color{blue} x_{1\hskip-0.15em} = 0, x_{3\hskip-0.09em} = 1$. Then, $\hat{\Phi} = (x_{2\hskip-0.07em} \vee \nx2) \wedge  [(\nx{2\hskip-0.05em} \wedge x_4) \vee (x_{2\hskip-0.05em} \wedge x_4) \vee  (\nx{2\hskip-0.05em} \wedge \nx4) \vee (x_{2\hskip-0.05em} \wedge \nx4)]$, constructed by removing  $\nx{1\hskip-0.13em}$ and $x_{3\hskip-0.09em}$ from $\Phi$ (see \cref{tQBF}). Note that $\Phi \gets  \nx{1\hskip-0.09em} \wedge x_{3\hskip-0.05em} \wedge (\scp^{_2\hskip-0.09em} \vee \scp^{_6}) \wedge (\scp^{_8\hskip-0.09em} \vee \scp^{_{11\!}} \vee \scp^{_{12\!}} \vee \scp^{_{14}})$.
\end{example}

\section{Conclusion}
Let $\Fm(\lt_j) \hskip-0.11em\coloneqq \lt_{j\hskip-0.07em} \wedge \Fm$. Then, $\lt_{j\hskip-0.07em} \wedge \Fm$ reduces to $\scp(\lt_j) \wedge \Fm'\hskip-0.05em(\lt_j)$ via X\.{O}R unless $\scp(\lt_j)$ is inconsistent such that $\scp(\lt_j)$ and $\Fm'\hskip-0.05em(\lt_j)$ are properly disjoint. That is, \color{red} if $\scp(\lt_j)$ is inconsistent, then $\unsat \Fm(\lt_j)$, \color{black} hence $\lt_{j\hskip-0.12em}$ is removed from $\Fm$ and $\scp \gets \scp \wedge \nlt_{j\hskip-0.05em}$. Thus, $\Fm$ reduces to $\scp \wedge \Fm'\hskip-0.09em$ unless $\scp$ is inconsistent. Also, $\scp$ and $\Fm'\hskip-0.09em$ are properly disjoint. Claim: $\Fm'\hskip-0.12em$ is satisfiable. Note that there is no difference in proving that $\Fm'\hskip-0.09em$ is satisfiable and proving that the inconsistency of the minterm $\scp(\lt_j)$ is necessary also for the unsatisfiability of the formula $\Fm(\lt_j)$. Proof sketch: $\scp \gets \scp \wedge \nlt_{j\hskip-0.12em}$ and $\ell \gets \ell \cup \{j\}$ if $\scp(\lt_j)$ is inconsistent, where $\ell = \Li - \bar{\ell}$.  Thus,  $\scp(\lt_i)$ is consistent for all $i \in \Li'\hskip-0.09em$  and each $\lt_{i\hskip-0.11em} \in X_{i\hskip-0.05em}$. As a result,  if  $i \notin (\Li'\hskip-0.07em \cup \bar{\ell})$, then  $\scp(\lt_i)$ is inconsistent. That is, if  $j \in \ell$, then $\scp(\lt_j)$ is inconsistent. Also, $j \in \ell$ iff $\unsat \Fm(\lt_j)$.  Therefore, \color{red} if $\unsat \Fm(\lt_j)$, then $\scp(\lt_j)$ is inconsistent.\color{black}

\bibliography{PvsNP}

\begin{thebibliography}{1}

\bibitem{Ros}
Kenneth~H. Rosen.
\newblock {\em Discrete Mathematics and its Applications}.
\newblock McGraw-Hill, seventh edition, 2012.

\bibitem{Sch78}
Thomas~J. Schaefer.
\newblock T{h}e complexity of satisfiability problems.
\newblock In {\em Proceedings of the Tenth Annual ACM Symposium on Theory of
  Computing}, STOC '78, pages 216--226, 1978.

\bibitem{Sip}
Michael Sipser.
\newblock {\em Introduction to the Theory of Computation}.
\newblock Cengage Learning, third edition, 2012.

\bibitem{Wig}
Avi Wigderson.
\newblock {\em Mathematics and Computation: A Theory Revolutionizing Technology
  and Science}.
\newblock Princeton University Press, 2019.

\end{thebibliography}

\appendix

\section{Graph Isomorphism}

Let $f\colon V \to \tilde{V}$ and $g\colon E \to \tilde{E}$. An isomorphism of  graphs $G$ and $\tilde{G}$ is a bijection between $V$ and $\tilde{V}$ such that any two vertices $v_{i\hskip-0.09em}$ and $v_{j\hskip-0.09em}$ are adjacent  iff $f(v_i)$ and $f(v_j)$ are adjacent.

\begin{definition}
The Boolean variable $x_{ij\hskip-0.15em}$ denotes that $v_{i\hskip-0.15em}$ is paired with $\tilde{v}_{j\hskip-0.05em}$, hence $v_{i\hskip-0.15em}$ and $\tilde{v}_{j\hskip-0.15em}$ are similar, viz., if $v_{i\hskip-0.09em} \leftrightarrow \tilde{v}_{j\hskip-0.05em}$, then $v_{i\hskip-0.09em} \sim \tilde{v}_{j\hskip-0.05em}$. Likewise, $y_{ij\hskip-0.15em}$ denotes that if $e_{i\hskip-0.09em} \leftrightarrow \tilde{e}_{j\hskip-0.05em}$, then $e_{i\hskip-0.09em} \sim \tilde{e}_{j\hskip-0.05em}$. Hence, $\nx{ij\hskip-0.15em}$ denotes that  $v_{i\hskip-0.09em} \nleftrightarrow \tilde{v}_{j\hskip-0.15em}$ and $\ny{ij\hskip-0.15em}$ denotes that $e_{i\hskip-0.09em} \nleftrightarrow \tilde{e}_{j\hskip-0.05em}$.
\end{definition}

Graph Isomorphism is tackled via an example (see \cref{f:GI}).

\begin{figure} [!h] \centering \vskip-0.4em
    \begin{tikzpicture}[transform shape, scale=0.9] \small

        \tikzset{label distance=-1.7mm}

            \node at (0,1.6)   [label=above:$v_3$] {\Bullet};
            \node at (1.6,1.6) [label=above:$v_4$] {\Bullet};

            \node at (0.8,0.8) [label=above:$v_5$] {\Bullet};

            \node at (0,0)     [label=below:$v_1$] {\Bullet};
            \node at (1.6,0)   [label=below:$v_2$] {\Bullet};

            \draw (0,0)     --  node[below] {$e_1$}            (1.6,0);   
            \draw (0,1.6)   --  node[above] {$e_3$}            (1.6,1.6); 
            \draw (0,0)     --  node[left]  {$G \colon \;\; e_4\!$} (0,1.6);   
            \draw (1.6,0)   --  node[right] {$\!e_2$}          (1.6,1.6); 
            \draw (0,0)     --  node[above] {$e_5\;$}          (0.8,0.8); 
            \draw (1.6,0)   --  node[above] {$\,e_6$}          (0.8,0.8); 
    \end{tikzpicture} \hskip7em
    \begin{tikzpicture}[transform shape, scale=0.9] \small

            \tikzset{label distance=-1.7mm}

            \node at (0,1.6)   [label=above:$\tilde{v}_1$] {\Bullet};
            \node at (1.6,1.6) [label=above:$\tilde{v}_3$] {\Bullet};

            \node at (0.8,0.8) [label=below:$\tilde{v}_2$] {\Bullet};

            \node at (0,0)     [label=below:$\tilde{v}_4$] {\Bullet};
            \node at (1.6,0)   [label=below:$\tilde{v}_5$] {\Bullet};

            \draw (0,0)     --  node[below] {$\tilde{e}_1$}                   (1.6,0);   
            \draw (0,1.6)   --  node[above] {$\tilde{e}_3$}                   (1.6,1.6); 
            \draw (0,0)     --  node[left] {$\tilde{G} \colon \;\; \tilde{e}_4\!$}         (0,1.6);   
            \draw (1.6,0)   --  node[right] {$\!\tilde{e}_2$}                 (1.6,1.6); 
            \draw (0,1.6)   --  node[below left] {$\tilde{e}_5\hskip-0.8em$}  (0.8,0.8); 
            \draw (1.6,1.6) --  node[below right] {$\hskip-0.6em \tilde{e}_6$} (0.8,0.8); 
    \end{tikzpicture}
  \caption{Graphs $G = (V, E)$ and $\tilde{G} = (\tilde{V}, \tilde{E})$.}\label{f:GI}
\end{figure}

Firstly,  $\FM_{\hskip-0.09em f\hskip-0.12em}$ is defined as follows with respect to the \emph{degrees} of the vertices in $V$ and $\tilde{V}$.

$\FM_{\hskip-0.09em f\hskip-0.12em} = (x_{11\hskip-0.11em} \X x_{13}) \wedge (x_{21\hskip-0.11em} \X x_{23}) \wedge (x_{32\hskip-0.05em} \X x_{34\hskip-0.05em} \X x_{35}) \wedge (x_{42\hskip-0.05em} \X x_{44\hskip-0.05em} \X x_{45}) \wedge (x_{52\hskip-0.05em} \X x_{54\hskip-0.05em} \X x_{55})$. Note that $x_{52\hskip-0.07em}$ denotes that $v_{5\hskip-0.09em} \leftrightarrow \tilde{v}_{2\hskip-0.05em}$. Note also that $v_{3\hskip-0.09em} \nsim \tilde{v}_{1\hskip-0.05em}$, i.e., $d(v_3) \neq d(\tilde{v}_1)$.

Because $f$ is a bijection, $f^{-1}$ can be defined by means of $\FM_{\hskip-0.09em f\hskip-0.09em}$. Then, $x_{11\hskip-0.11em} \imp \nx{21\hskip-0.05em}$, $x_{13\hskip-0.11em} \imp \nx{23\hskip-0.05em}$, $x_{32\hskip-0.11em} \imp \nx{42\hskip-0.05em} \wedge \nx{52\hskip-0.05em}$, $x_{34\hskip-0.11em} \imp \nx{44\hskip-0.05em} \wedge \nx{54\hskip-0.05em}$, $x_{35\hskip-0.11em} \imp \nx{45\hskip-0.05em} \wedge \nx{55}$, $x_{42\hskip-0.11em} \imp \nx{52\hskip-0.05em}$, $x_{44\hskip-0.11em} \imp \nx{54\hskip-0.05em}$, and $x_{45\hskip-0.11em} \imp  \nx{55}$. That is, $\FM_{\hskip-0.09em f^{\hskip-0.05em*}\hskip-0.11em} = (\nx{11\hskip-0.21em} \vee \nx{21}) \wedge (\nx{13\hskip-0.21em} \vee \nx{23}) \wedge (\nx{32\hskip-0.21em} \vee \nx{42}) \wedge (\nx{32\hskip-0.21em} \vee \nx{52}) \wedge (\nx{34\hskip-0.21em} \vee \nx{44}) \wedge (\nx{34\hskip-0.21em} \vee \nx{54}) \wedge (\nx{35\hskip-0.21em} \vee \nx{45}) \wedge (\nx{35\hskip-0.21em} \vee \nx{55}) \wedge (\nx{42\hskip-0.21em} \vee \nx{52}) \wedge (\nx{44\hskip-0.21em} \vee \nx{54}) \wedge (\nx{45\hskip-0.21em} \vee \nx{55})$.

Next, $\FM_{\hskip-0.09em g\hskip-0.12em}$ is defined as follows with respect to the degrees of the vertices in $V$ and $\tilde{V}$.

$\FM_{\hskip-0.09em g\hskip-0.12em} = \color{red}y_{13\hskip-0.07em} \wedge y_{31\hskip-0.09em} \color{black}\wedge (y_{22\hskip-0.07em} \X y_{24\hskip-0.11em} \X y_{25\hskip-0.05em} \X y_{26}) \wedge (y_{42\hskip-0.07em} \X y_{44\hskip-0.11em} \X y_{45\hskip-0.05em} \X y_{46}) \wedge (y_{52\hskip-0.07em} \X y_{54\hskip-0.11em} \X y_{55\hskip-0.05em} \X y_{56}) \wedge (y_{62\hskip-0.07em} \X y_{64\hskip-0.11em} \X y_{65\hskip-0.05em} \X y_{66})$. Note that $y_{13\hskip-0.07em}$ denotes that $e_{1\hskip-0.19em} \leftrightarrow \tilde{e}_{3\hskip-0.03em}$. Note also that $e_{1\hskip-0.19em} \nsim \tilde{e}_{1\hskip-0.05em}$, because the degrees of $\{v_1, v_2\}$ are not paired with the degrees of $\{\tilde{v}_4, \tilde{v}_5\}$, that is, $d(v_1) \neq d(\tilde{v}_4)$ and $d(v_1) \neq d(\tilde{v}_5)$, and $d(v_2) \neq d(\tilde{v}_4)$ and $d(v_2) \neq d(\tilde{v}_5)$. Likewise, $e_{1\hskip-0.19em} \nsim \tilde{e}_{2\hskip-0.05em}$.

Because $g$ is a bijection, $g^{-1}$ can be defined by means of $\FM_{\hskip-0.09em g\hskip-0.09em}$. Then, $y_{2j\hskip-0.11em} \imp \ny{4j\hskip-0.09em} \wedge \ny{5j\hskip-0.09em} \wedge \ny{6j\hskip-0.05em}$, $y_{4j\hskip-0.11em} \imp \ny{5j\hskip-0.09em} \wedge \ny{6j\hskip-0.05em}$, and $y_{5j\hskip-0.11em} \imp \ny{6j\hskip-0.09em}$ for any $j \in \{2, 4, 5, 6\}$. That is, $\FM_{\hskip-0.09em g^{\hskip-0.05em*}\hskip-0.11em} = \bigwedge_{j \in \{2, 4, 5, 6\}}(\ny{2j\hskip-0.21em} \vee \ny{4j}) \wedge (\ny{2j\hskip-0.21em} \vee \ny{5j})  \wedge (\ny{2j\hskip-0.21em} \vee \ny{6j}) \wedge (\ny{4j\hskip-0.21em} \vee \ny{5j}) \wedge (\ny{4j\hskip-0.21em} \vee \ny{6j})  \wedge (\ny{5j\hskip-0.21em} \vee \ny{6j})$.

Finally, $x_{ij\hskip-0.12em}$ and $y_{ij\hskip-0.12em}$ are related by means of $\FM_{x_{ij\hskip-0.12em}}$ and $\FM_{y_{ij\hskip-0.05em}}$, some of which are specified below. Note that $\FM_{x_{ij\hskip-0.15em}}$ is defined over $\FM_{\hskip-0.09em f\hskip-0.09em}$, and $\FM_{y_{ij\hskip-0.15em}}$ is defined over $\FM_{\hskip-0.09em g\hskip-0.09em}$. For example, if $\color{blue}e_{2\hskip-0.09em} \leftrightarrow \tilde{e}_{2\hskip-0.05em}$, then $\color{blue}v_{2\hskip-0.09em} \leftrightarrow \tilde{v}_{3\hskip-0.07em}$ and $\color{blue}v_{4\hskip-0.09em} \leftrightarrow \tilde{v}_{5\hskip-0.05em}$. Note that $d(v_4) = d(\tilde{v}_5)$ and $d(v_2) = d(\tilde{v}_3)$, where $d(v_4) = 2$ and $d(v_2) = 3$. Also, if $v_{5\hskip-0.09em} \leftrightarrow \tilde{v}_{5\hskip-0.05em}$, then  $e_{5\hskip-0.09em} \leftrightarrow \tilde{e}_{2\hskip-0.09em}$ or $e_{6\hskip-0.09em} \leftrightarrow \tilde{e}_{2\hskip-0.03em}$.

\noindent $\FM_{x_{11\hskip-0.21em}} = \nx{11\hskip-0.15em} \vee y_{13\hskip-0.15em} \vee y_{44\hskip-0.15em} \vee y_{45\hskip-0.15em}\vee  y_{54\hskip-0.15em} \vee y_{55\hskip-0.05em}$. \tabto{24em} $\FM_{y_{13\hskip-0.15em}} = \ny{13\hskip-0.15em} \vee x_{11\hskip-0.15em} \vee x_{13\hskip-0.15em} \vee x_{21\hskip-0.15em} \vee x_{23\hskip-0.05em}$.

\noindent $\FM_{x_{13\hskip-0.15em}} = \nx{13\hskip-0.15em} \vee y_{13\hskip-0.15em} \vee y_{42\hskip-0.15em} \vee y_{46\hskip-0.15em} \vee  y_{52\hskip-0.15em} \vee y_{56\hskip-0.05em}$. \tabto{24em} $\FM_{y_{31\hskip-0.15em}} = \ny{31\hskip-0.15em} \vee x_{34\hskip-0.15em} \vee x_{35\hskip-0.15em} \vee x_{44\hskip-0.15em} \vee x_{45\hskip-0.05em}$.

\noindent $\FM_{x_{21\hskip-0.21em}} = \nx{21\hskip-0.15em} \vee y_{13\hskip-0.15em} \vee y_{24\hskip-0.15em} \vee y_{25\hskip-0.15em} \vee  y_{64\hskip-0.15em} \vee y_{65\hskip-0.05em}$. \tabto{24em} $\color{blue}\FM_{y_{22\hskip-0.15em}} = (\ny{22\hskip-0.15em} \vee x_{23}) \wedge (\ny{22\hskip-0.15em} \vee x_{45})$.

\noindent $\FM_{x_{55\hskip-0.15em}} = \nx{55\hskip-0.15em} \vee y_{52\hskip-0.15em} \vee  y_{62\hskip-0.05em}$. \tabto{24em} $\FM_{y_{66\hskip-0.15em}} = (\ny{66\hskip-0.11em} \vee x_{23}) \wedge (\ny{66\hskip-0.11em} \vee x_{52})$.

Let $\FM = \FM_{\hskip-0.09em f\hskip-0.09em} \wedge \FM_{\hskip-0.09em f^{\hskip-0.05em*}\hskip-0.19em} \wedge \FM_{\hskip-0.09em g\hskip-0.09em} \wedge \FM_{\hskip-0.09em g^{\hskip-0.05em*}\hskip-0.19em} \wedge \FM_{x_{11\hskip-0.17em}} \wedge \FM_{x_{13\hskip-0.11em}} \wedge \cdots \wedge \FM_{x_{55\hskip-0.11em}} \wedge \FM_{y_{13\hskip-0.11em}} \wedge \FM_{y_{31\hskip-0.17em}} \wedge \cdots \wedge \FM_{y_{66\hskip-0.07em}}$, which denotes an XSAT formula, after $\FM_{\hskip-0.09em f^{\hskip-0.05em*}\hskip-0.09em}$, $\FM_{\hskip-0.09em g^{\hskip-0.05em*}\hskip-0.09em}$, $\FM_{x_{ij\hskip-0.12em}}$ and $\FM_{y_{ij\hskip-0.12em}}$ are transformed into an X3SAT formula. As a result, $G$ and $\tilde{G}$ are isomorphic iff $\FM$ is satisfiable. Therefore, a satisfying assignment (see \cref{s:Assg}) denotes an isomorphism. Note that $\FM = \scp \wedge \Fm$, where $\color{red}\scp = y_{13\hskip-0.07em} \wedge y_{31\hskip-0.09em}$. Note also that $\Fm =  \FM_{\hskip-0.09em f\hskip-0.09em} \wedge \FM_{\hskip-0.09em f^{\hskip-0.05em*}\hskip-0.19em} \wedge (y_{22\hskip-0.07em} \X y_{24\hskip-0.11em} \X y_{25\hskip-0.05em} \X y_{26}) \wedge (y_{42\hskip-0.07em} \X y_{44\hskip-0.11em} \X y_{45\hskip-0.05em} \X y_{46}) \wedge (y_{52\hskip-0.07em} \X y_{54\hskip-0.11em} \X y_{55\hskip-0.05em} \X y_{56}) \wedge (y_{62\hskip-0.07em} \X y_{64\hskip-0.11em} \X y_{65\hskip-0.05em} \X y_{66}) \wedge \FM_{\hskip-0.09em g^{\hskip-0.05em*}\hskip-0.19em} \wedge \FM_{x_{11\hskip-0.15em}} \wedge \FM_{x_{13\hskip-0.11em}} \wedge \cdots \wedge \FM_{x_{55\hskip-0.1em}} \wedge \FM_{y_{13}\hskip-0.07em}(\neg \ny{13}) \wedge \FM_{y_{31}\hskip-0.09em}(\neg \ny{31}) \wedge \cdots \wedge \FM_{y_{66\hskip-0.05em}}$, in which $\FM_{y_{13}\hskip-0.07em}(\neg \ny{13}) = x_{11\hskip-0.15em} \vee x_{13\hskip-0.15em} \vee x_{21\hskip-0.15em} \vee x_{23\hskip-0.03em}$. Note that $y_{13\hskip-0.11em}$ is necessary for $\FM$, i.e., $\color{red}y_{13\hskip-0.07em} \in \scp$. Next, incompatibility of $x_{ij\hskip-0.12em}$ and $y_{ij\hskip-0.12em}$ are checked  by means of \cref{noScope}.

\end{document}